
\magnification=\magstep1

\font\eighteenrm = cmr10 scaled\magstep3
\font\eighteeni = cmmi10 scaled\magstep3
\font\eighteensy = cmsy10 scaled\magstep3
\font\eighteenit = cmti10 scaled\magstep3
\font\eighteenb = cmbx10 scaled\magstep3

\font\twelverm = cmr12

\font\twelvei = cmmi12
\font\twelveit = cmti12
\font\twelveb = cmbx12
\font\twelvesy = cmsy10 scaled\magstep1
\font\twelves = cmsl12

\font\tenrm = cmr10
\font\tenit = cmti10
\font\teni = cmmi10                  
\font\tensy = cmsy10                
\font\tens = cmsl10                  
\font\tenb = cmbx10                  

\font\teniu = cmu10
\font\ninerm = cmr9
\font\ninesy = cmsy9

\font\eightrm = cmr8
\font\eighti = cmmi8
\font\eightsy = cmsy8

\font\sixrm = cmr6
\font\sixi  = cmmi6
\font\sixsy = cmsy6

\font\fivesy = cmsy5

\def\tenpoint{\def\rm{\fam0\tenrm}
\textfont0=\tenrm \scriptfont0=\eightrm \scriptscriptfont0=\sixrm
\textfont1=\teni \scriptfont1=\eighti \scriptscriptfont1=\sixi
\textfont2=\tensy \scriptfont2=\eightsy \scriptscriptfont2=\sixsy
\textfont3=\tenex \scriptfont3=\tenex \scriptscriptfont3=\tenex
\def\sy{\fam4\tensy}%
\textfont4=\tensy%
\def\sl{\fam5\tens}%
\textfont5=\tens%
\def\bf{\fam6\tenb}%
\textfont6=\tenb%
\def\it{\fam7\tenit}%
\textfont7=\tenit%
\def\prfnt{\fam8\ninesy}%
\textfont8=\ninesy%
\def\hbfnt{\fam9\fivesy}%
\textfont9=\fivesy%
\textfont11=\sixrm \scriptfont11=\sixrm \scriptscriptfont11=\sixrm%
\baselineskip 12pt                                                
\lineskip 1pt
\parskip 5pt plus 1pt
\abovedisplayskip 12pt plus 3pt minus 9pt
\belowdisplayshortskip 7pt plus 3pt minus 4pt \tenrm}
\def\prfnt{\ninesy }
\def\hbfnt{\fivesy }

\def\bigfnt{\twelverm}

\def\eighteenpoint{\def\rm{\fam0\eighteenrm}%
\textfont0=\eighteenrm \scriptfont0=\twelverm \scriptscriptfont0=\eightrm
\textfont1=\eighteeni \scriptfont1=\twelvei \scriptscriptfont1=\eighti
\textfont2=\eighteensy \scriptfont2=\twelvesy \scriptscriptfont2=\eightsy
\textfont3=\tenex \scriptfont3=\tenex \scriptscriptfont3=\tenex
\def\sy{\fam4\eighteensy}%
\textfont4=\eighteensy%
\def\bf{\fam6\eighteenb}%
\textfont6=\eighteenb%
\def\it{\fam7\eighteenit}%
\textfont7=\eighteenit%
\baselineskip 21pt
\lineskip 1pt
\parskip 5pt plus 1pt
\abovedisplayskip 15pt plus 5pt minus 10pt
\belowdisplayskip 15pt plus 5pt minus 10pt
\abovedisplayshortskip 13pt plus 8pt
\belowdisplayshortskip 10pt plus 5pt minus 5pt
\eighteenrm}
\def\twelvepoint{\def\rm{\fam0\twelverm}%
\textfont0=\twelverm \scriptfont0=\tenrm \scriptscriptfont0=\eightrm
\textfont1=\twelvei \scriptfont1=\teni \scriptscriptfont1=\eighti
\textfont2=\twelvesy \scriptfont2=\tensy \scriptscriptfont2=\eightsy
\textfont3=\tenex \scriptfont3=\tenex \scriptscriptfont3=\tenex
\def\sy{\fam4\twelvesy}%
\textfont4=\twelvesy%
\def\sl{\fam5\twelves}%
\textfont5=\twelves%
\def\bf{\fam6\twelveb}%
\textfont6=\twelveb%
\def\it{\fam7\twelveit}%
\textfont7=\twelveit%
\def\prfnt{\fam8\ninesy}%
\textfont8=\ninesy%
\def\hbfnt{\fam9\fivesy}%
\textfont9=\fivesy%
\def\paperfont{\fam10\twelverm}%
\def\dotfont{\fam11\sixrm}%
\textfont11=\sixrm \scriptfont11=\sixrm \scriptscriptfont11=\sixrm%
\baselineskip 15pt
\lineskip 1pt
\parskip 5pt plus 1pt
\abovedisplayskip 15pt plus 5pt minus 10pt
\belowdisplayskip 15pt plus 5pt minus 10pt
\abovedisplayshortskip 13pt plus 8pt
\belowdisplayshortskip 10pt plus 5pt minus 5pt \twelverm}

\tenpoint
\hsize 5.9truein
\vsize 9.5truein
\global \topskip .7truein
\newcount\chapnum
\chapnum = 0
\newcount\sectnum
\sectnum = 0
\newcount\subsecnum            
\subsecnum = 0        
\newcount\eqnum
\newcount\chapnum
\chapnum = 0
\newcount\sectnum
\sectnum = 0
\newcount\subsecnum            
\subsecnum = 0        
\newcount\eqnum
\eqnum = 0
\newcount\refnum
\refnum = 0
\newcount\tabnum
\tabnum = 0
\newcount\fignum
\fignum = 0
\newcount\footnum
\footnum = 1
\newcount\pointnum
\pointnum = 0
\newcount\subpointnum
\subpointnum = 96
\newcount\subsubpointnum
\subsubpointnum = -1
\newcount\letnum
\letnum = 0
\newbox\referens
\newbox\figures
\newbox\tables
\newbox\tempa
\newbox\tempb
\newbox\tempc
\newbox\tempd
\newbox\tempe
\hbadness=10000
\newbox\refsize
\setbox\refsize\hbox to\hsize{ }
\hbadness=1000
\newskip\refbetweenskip
\refbetweenskip = 5pt
\def\ctrline#1{\line{\hss#1\hss}}
\def\rjustline#1{\line{\hss#1}}
\def\ljustline#1{\line{#1\hss}}

\def\ctr#1{\hfill{#1}\hfill}

\def\spose#1{\hbox to 0pt{#1\hss}}
\def\chskipt{\vskip .125in plus 0pt minus 0pt }              
\def\chskipl{\vskip .7in plus 18pt minus 10pt}               
\def\secskipt{\penalty-500\vskip 24pt plus 2pt minus 2pt}    
\def\secskipl{\vskip 3.5pt plus 1pt }
\def\subsecskip{\penalty-500\vskip 6pt plus 2pt minus 2pt }
\def\unchskip{\vskip -.7in }                                 
\def\conskip{\vskip 14pt }
\newif\ifoddeven
\gdef\oneside{\oddevenfalse}

\oneside
\newif\ifnonumpageone

\gdef\nonumberfirst{\nonumpageonetrue}
\nonumberfirst
\output{\ifoddeven\leftright\else\samemarg\fi
        \ifnonumpageone\checkpage\else\empty\fi
        \plainoutput}
\def\leftright{\ifodd\count0{\global\hoffset=\oddmargin}
                       \else{\global\hoffset=\evenmargin}\fi}
\def\samemarg{\global\hoffset=\oddmargin}
\gdef\oddmargin{.25truein}
\gdef\evenmargin{0truein}
\def\checkpage{\ifnum\count0=1\nopagenumbers\else\empty\fi}
\footline={{\pagefont\hss--\hquad\folio\hquad--\hss}}
\def\pagefont{\teniu}
\setbox\referens\vbox{\ctrline{\bf References }\chskipt }
\setbox\figures\vbox{\ctrline{\bf Figure Captions }\chskipt }
\setbox\tables\vbox{\ctrline{\bf Table Captions }\chskipt }

\def\title#1{\ctrline {\titfnt #1} }
\def\titfnt{\eighteenpoint}
\def\author#1{\ctrline{\autfnt #1}\par}
\def\autfnt{\bigfnt}

\def\abstract{
\ctrline{\bf ABSTRACT}\chskipt}

\def\reset{\global\sectnum = 0 \global\eqnum = 0
     \global\subsecnum = 0}
\def\chap#1{\global\advance\chapnum by 1 \reset 
\endpage
\chskipt
\ifodd\count0{
\rightline{{\chapnumfont Chapter \the\chapnum}}
\medskip
\rightline{{\chapfont #1}}}
\else{
\leftline{{\chapnumfont Chapter \the\chapnum}}
\medskip
\ljustline{{\chapfont #1}}}\fi
\penalty 100000 \chskipl  \penalty 100000
{\let\number=0\edef\next{
\write2{\bigskip\noindent
  \tofcfont Chapter \the\chapnum.{ }#1
  \leadtofc\number\count0\smallskip}}
\next}}
\def\titcon#1{\unchskip
\ifodd\count0{
\rightline{{\chapfont #1}}}
\else{
\leftline{{\chapfont #1}}}\fi
\penalty 100000 \chskipl \penalty 100000 }
\def\chapnumfont{\tenit}
\def\chapfont{\eighteenpoint}
\def\chapter#1{\chap{#1}}
\def\sect#1{\global\advance\sectnum by 1 \global\subsecnum = 0
\secskipt
\ifnum\chapnum=0{{\sectfont\par\noindent
\secsign\the\sectnum{ }{ }#1}\par
         {\let\number=0\edef\next{
         \write2{\vskip 0pt\noindent\hangindent 30pt
         \tofcfont\hbox to 30pt{\hfill\the\sectnum\quad}\unskip#1
         \leadtofc\number\count0}}
         \next}}
\else{{\sectfont\par\noindent
\secsign\the\chapnum .\the\sectnum{ }{ }#1}\par
      {\let\number=0\edef\next{
       \write2{\vskip 0pt\noindent\hangindent 30pt
       \tofcfont\hbox to 30pt{\hfill\the\chapnum
              .\the\sectnum\quad}\unskip#1
         \leadtofc\number\count0}}
       \next}}\fi
       \nobreak\medskip\nobreak}
\def\sectfont{\twelvepoint}
\def\secsign{\S}

\def\subsect#1{\global\advance\subsecnum by 1 \secskipt
\noindent
\ifnum\chapnum=0{
     {\subsecfont\the\sectnum .\the\subsecnum{ }{ }#1}
      \nobreak\par
     {\let\number=0\edef\next{
     \write2{\vskip 0pt\noindent\hangindent 58pt\tofcfont
     \hbox to 60pt{\hfill\the\sectnum
           .\the\subsecnum\quad}\unskip#1
     \leadtofc\number\count0}}\next}}
\else{{
     \subsecfont\the\chapnum .\the\sectnum
        .\the\subsecnum{ }{ }#1}\nobreak\par
     {\let\number=0\edef\next{
     \write2{\vskip 0pt\noindent\hangindent 58pt\tofcfont
     \hbox to 60pt{\hfill\the\chapnum.\the\sectnum
           .\the\subsecnum\quad}\unskip#1
     \leadtofc\number\count0}}\next}}\fi
     \nobreak\medskip\nobreak}

\def\subsecfont{\twelvepoint}        
\immediate\openout 2 tofc
\def\tableofcontents#1{\endpage
\count0=#1
\chskipt
\ifodd0{
\rjustline{{\chapfont Contents}}}
\else{
\ljustline{{\chapfont Contents}}}\fi
\chskipl
\rjustline{{\tofcfont Page}}
\bigskip
\immediate\closeout 2
\input tofc
\endpage}

\def\tofcfont{\ninerm}
\def\leadtofc{\leaders\hbox to 8pt{\hfill.\hfill}\hfill}
\immediate\openout 4 refc
\def\refbegin#1#2{\unskip\global\advance\refnum by1
\xdef\rnum{\the\refnum}
\xdef#1{\the\refnum}
\xdef\rtemp{$^{\rnum}$}
\unskip
\immediate\write4{\vskip 5pt\par\noindent\tofcfont
  \hangindent .11\wd\refsize \hbox to .11\wd\refsize{\hfill
  \the\refnum . \quad } \unskip}\unskip
  \immediate\write4{#2}\unskip}
\def\refend{\nobreak\rtemp\unskip}
\def\ref#1{\refbegin{\?}{#1}}

\def\REF#1#2{\refbegin{#1}{#2}}
\def\refsbegin#1#2{\unskip\global\advance\refnum by 1
\xdef\refb{\the\refnum}
\xdef#1{\the\refnum}
\xdef\rrnum{\the\refnum}
\unskip
\immediate\write4{\vskip 5pt\par\noindent\tofcfont
  \hangindent .11\wd\refsize \hbox to .11\wd\refsize{\hfill
  \the\refnum . \quad } \unskip}\unskip
  \immediate\write4{#2}\unskip}
\def\REFSCON#1#2{\unskip \global\advance\refnum by 1
\xdef#1{\the\refnum}
\xdef\rrrnum{\the\refnum}
\unskip
\immediate\write4{\vskip 5pt\par\noindent\tofcfont
  \hangindent .11\wd\refsize \hbox to .11\wd\refsize{\hfill
  \the\refnum . \quad } \unskip}\unskip
  \immediate\write4{#2}\unskip}

\def\refsend{\nobreak$^{\refb-\the\refnum}$\unskip}

\def\REFS#1#2{\refsbegin{#1}{#2} }
\def\foot#1{\footnote{$^{\the\footnum}$}{#1}
  \global\advance\footnum by 1}

\def\figure#1#2{\global\advance\fignum by 1
\xdef#1{\the\fignum }
\ctrline{\Figure . #2}\par\conskip
{\let\number=0\edef\next{
\write3{\par\noindent\tofcfont
  \hangindent .11\wd\refsize \hbox to .11\wd\refsize{\hfill
  \the\fignum . \quad } \unskip}
\write3{#2\leadtofc\number\count0\par}}
\next}}
\def\figurs#1#2#3{\global\advance\fignum by 1
\xdef#1{\the\fignum }
\ctrline{\Figure . \it #2}
\ctrline{\it #3}\par\conskip
{\let\number=0\edef\next{
\write3{\par\noindent\tofcfont
  \hangindent .11\wd\refsize \hbox to .11\wd\refsize{\hfill
  \the\fignum . \quad}\unskip}
\write3{#2 #3\leadtofc\number\count0\par}}
\next}}

\def\figcon{\ctrline{{\it Figure  \the\fignum} -- cont'd}\par\conskip}
\def\Figure{{\it Figure  \the\fignum}}
\immediate\openout 3 figc

\def\table#1#2{\global\advance\tabnum by 1
\xdef#1{\the\tabnum }
\ctrline{\Table . #2}\par\conskip
{\let\number=0\edef\next{
\write5{\par\noindent\tofcfont
  \hangindent .11\wd\refsize \hbox to .11\wd\refsize{\hfill
  \the\tabnum . \quad } \unskip}
\write5{#2\leadtofc\number\count0\par}}
\next}}
\def\tabls#1#2#3{\global\advance\tabnum by 1
\xdef#1{\the\tabnum }
\ctrline{\Table . #2}
\ctrline{\it #3}\par\conskip
{\let\number=0\edef\next{
\write5{\par\noindent\tofcfont
  \hangindent .11\wd\refsize \hbox to .11\wd\refsize{\hfill
  \the\tabnum . \quad}\unskip}
\write5{#2 #3\leadtofc\number\count0\par}}
\next}}

\def\Table{\it Table  \the\tabnum}
\immediate\openout 5 tabc

\def\eqname#1{\global\advance\eqnum by 1
\ifnum\chapnum=0{
   \xdef#1{ (\the\eqnum ) }(\the\eqnum )  }
\else{
   \xdef#1{ (\the\chapnum .\the\eqnum ) }
            (\the\chapnum .\the\eqnum ) }\fi}
\def\enum{\global\advance\eqnum by 1
  \ifnum\chapnum=0{ (\the\eqnum )  }
  \else{(\the\chapnum .\the\eqnum ) }\fi}

\def\eqn#1{\eqno \eqname{#1} }
\def\eqnameap#1{\global\advance\eqnum by 1
   \xdef#1{ (\copy\appbox .\the\eqnum ) }
            (\copy\appbox .\the\eqnum ) }
\def\enumap{\global\advance\eqnum by 1
  (\copy\appbox .\the\eqnum ) }

\def\item#1{\par\noindent\hangindent .08\wd\refsize
\hbox to .08\wd\refsize{\hfill #1\quad}\unskip}
\def\sitem#1{\par \noindent\hangindent .13\wd\refsize
\hbox to .13\wd\refsize{\hfill #1\quad}\unskip}
\def\ssitem#1{\par\noindent\hangindent .195\wd\refsize
\hbox to .195\wd\refsize{\hfill #1\quad}\unskip}

\def\point{\par \global\advance\pointnum by 1
\noindent\hangindent .08\wd\refsize \hbox to .08\wd\refsize{\hfill
\the\pointnum .\quad}\unskip}

\def\spoint{\par \global\advance\subpointnum by 1
\noindent\hangindent .13\wd\refsize
\hbox to .13\wd\refsize{\hfill
(\char\the\subpointnum )\quad}\unskip}

\def\sspoint{\par \global\advance\subsubpointnum by 1
\noindent\hangindent .195\wd\refsize
\hbox to .195\wd\refsize {\hfill\hbox to 20pt{
(\romannumeral\subsubpointnum\hfill)}\quad}\unskip}

\def\bye{\endpage\end}              

\def\bspace#1{\hbox to -#1{}}
\newbox\appbox
\def\appendix#1{\endpage\reset
\setbox\appbox\hbox{#1}
\chskipt \ctrline {\bf APPENDIX #1 }\penalty 10000
\chskipl \penalty 10000
\write2{\bigskip\noindent
 {\tofcfont Appendix #1\leadtofc\number\count0\par\smallskip}}}

\def\mat#1#2{\if 2#1 {\left( \  \vcenter{\halign{$\ctr{## }$ \quad
& $\ctr{## }$\cr #2}} \  \right) } \else{ }\fi
\if 3#1 {\left( \  \vcenter{\halign{
$\ctr{## }$ \quad & $\ctr{## }$ \quad
& $\ctr{## }$\cr #2}} \  \right) } \else{ }\fi
\if 4#1 {\left( \  \vcenter{\halign{$\ctr{## }$ \quad &
$\ctr{## }$ \quad & $\ctr{## }$ \quad
& $\ctr{## }$\cr #2}} \  \right) } \else{ }\fi
\if 5#1 {\left( \  \vcenter{\halign{$\ctr{## }$ \quad
& $\ctr{## }$ \quad & $\ctr{## }$ \quad
& $\ctr{## }$ \quad & $\ctr{## }$\cr #2}} \  \right)} \else{ }\fi
\if 6#1 {\left( \  \vcenter{\halign{$\ctr{## }$ \quad
& $\ctr{## }$ \quad & $\ctr{## }$ \quad & $\ctr{## }$ \quad
& $\ctr{## }$ \quad & $\ctr{## }$\cr #2}} \  \right)} \else{ }\fi }

\def\endpage{\par \vfill \eject}
\def\physrev{\baselineskip 24pt
\lineskip 1pt
\parskip 1pt plus 1pt
\abovedisplayskip 15pt plus 7pt minus 13.33pt
\belowdisplayskip 15pt plus 7pt minus 13.33pt
\abovedisplayshortskip 14pt plus 11pt
\belowdisplayshortskip 9pt plus 7pt minus 7pt
\def\chskipt{\vskip 24pt }
\def\chskipl{\vskip 6.5pt }
\def\secskipt{\vskip 7pt plus 3pt minus 1.33pt }
\def\secskipl{\vskip 3.5pt plus 2pt }
\def\subsecskip{\vskip 7pt plus 2pt minus 2pt }
\def\unchskip{\vskip -6.5pt }
\def\conskip{\vskip 24pt }
\refbetweenskip = \the\baselineskip
\multskip\refbetweenskip by 5
\divskip\refbetweenskip by 10
\twelverm }
\def\bk{\hfil\break}         

\def\To{\par\noindent\hangindent .18\wd\refsize
\hbox to .18\wd\refsize {To: \hfill \qquad }}
\def\from{\par\noindent\hangindent .18\wd\refsize
\hbox to .18\wd\refsize {From: \hfill \qquad }}
\def\topic{\par\noindent\hangindent .18\wd\refsize
\hbox to .18\wd\refsize{Topic: \hfill \qquad }}

\def\startpage#1 {\count0 = #1}
\def\startchapter#1{\chapnum = #1 \advance\chapnum by -1}

\def\startfig#1{\fignum = #1 \advance\fignum by -1}
\def\starttab#1{\tabnum = #1 \advance\tabnum by -1}

\def\rarrow{\rightarrow }

\newbox\xa\newbox\xb
\def\boxit#1{ \setbox\xa\vbox {\vskip \boxitsep
\hbox{\hskip \boxitsep #1\hskip \boxitsep }\vskip \boxitsep }
\setbox\xb\hbox{\vrule \copy\xa \vrule}
\vbox{\hrule width 1\wd\xb \copy\xb \hrule width 1\wd\xb }}
\def\fboxit#1#2{ \setbox\xa\vbox {\vskip \boxitsep
\hbox{\hskip \boxitsep #2\hskip \boxitsep }\vskip \boxitsep }
\setbox\xb\hbox{\vrule width #1pt \copy\xa \vrule width #1pt}
\vbox{\hrule height #1pt width 1\wd\xb
\copy\xb \hrule height #1pt width 1\wd\xb }}
\def\reboxit#1#2#3{ \setbox\xa\vbox{\vskip \boxitsep
\hbox{\hskip \boxitsep #3\hskip \boxitsep }\vskip \boxitsep }
\setbox\xb\hbox{\vrule width #1pt\bspace{#2}
\copy\xa \vrule width #1pt}
\vbox{\hrule height #1pt width 1\wd\xb
\copy\xb \hrule height #1pt width 1\wd\xb}}
\def\boxitsep{4pt}

\newdimen\offdimen
\def\offset#1#2{\offdimen #1
   \noindent \hangindent \offdimen
   \hbox to \offdimen{#2\hfil}\ignorespaces}
\newdimen\defnamespace   
\defnamespace=2in        
\def\definition#1#2{     
    \def\itema{\par\hang\textindent}
    {\advance\parindent by \defnamespace
     \advance\defnamespace by -.5em
     \itema{\hbox to \defnamespace{#1\hfil}}#2\par}}
\def\TEX{\hbox{T\hskip-2pt\lower1.94pt\hbox{E}\hskip-2pt X}}
\def\wyz{\hbox{WI\hskip-1pt\lower.9pt\hbox{Z\hskip-1.85pt
\raise1.7pt\hbox{Z}}LE}}
\def\\{$\backslash $}
\def\rc{$\} $}
\def\underwiggle#1{\mathop{\vtop{\ialign{##\crcr
    $\hfil\displaystyle{#1}\hfil$\crcr\noalign{\kern2pt\nointerlineskip}
    $\scriptscriptstyle\sim$\crcr\noalign{\kern2pt}}}}\limits}
\def\({[}
\def\){]}

\def\hquad{\hskip.5em{}}
\mathchardef\app"3218

\def\linebreak{\break}
\mathchardef\oprod="220A
\mathchardef\inter="225C
\mathchardef\union="225B

\mathchardef\relv="326A
\mathchardef\leftv="326A
\mathchardef\rightv="326A
\mathchardef\relvv"326B
\mathchardef\leftvv"326B
\mathchardef\rightvv"326B
\mathchardef\Zscr"25A
\mathchardef\Yscr"259
\mathchardef\Xscr"258
\mathchardef\Wscr"257
\mathchardef\Vscr"256
\mathchardef\Uscr"255
\mathchardef\Tscr"254
\mathchardef\Sscr"253
\mathchardef\Rscr"252
\mathchardef\Qscr"251
\mathchardef\Pscr"250
\mathchardef\Oscr"24F
\mathchardef\Nscr"24E
\mathchardef\Mscr"24D
\mathchardef\Lscr"24C
\mathchardef\Kscr"24B
\mathchardef\Jscr"24A
\mathchardef\Iscr"249
\mathchardef\Hscr"248
\mathchardef\Gscr"247
\mathchardef\Fscr"246
\mathchardef\Escr"245
\mathchardef\Dscr"244
\mathchardef\Cscr"243
\mathchardef\Bscr"242
\mathchardef\Ascr"241
\mathchardef\lscr"160

\immediate\openout 2 tofc
\immediate\openout 3 figc
\immediate\openout 4 refc
\immediate\openout 5 tabc
\tolerance 4000
\def\refbegin#1#2{\unskip\global\advance\refnum by1
\xdef\rnum{\the\refnum}
\xdef#1{\the\refnum}
\xdef\rtemp{[\rnum]}
\unskip
\immediate\write4{\vskip 5pt\par\noindent\tofcfont
  \hangindent .11\wd\refsize \hbox to .11\wd\refsize{\hfill
  \the\refnum . \quad } \unskip}\unskip
  \immediate\write4{#2}\unskip}
\def\refsend{\nobreak[\refb-\the\refnum]\unskip}

\def\refesh#1{\(#1\)}
\def\({\lbrack}
\def\){\rbrack}
\def\m{\mu}
\def\l{\lambda}
\def\a{\alpha}
\def\b{\beta}
\def\c{\gamma}
\def\d{\delta}
\def\n{\nu}
\def\om{\omega}
\def\deriv#1#2{{d#1\over d#2}}
\def\pd#1#2{{\partial#1\over \partial#2}}
\def\div{\vec\nabla\cdot}

\def\grad{\vec\nabla}
\def\sss{\scriptscriptstyle}
\def\&#1{^{\sss #1}}
\def\_#1{_{\sss #1}}
\def\bk{\par\noindent}
\def\abunit{km/s/Mpc}
\def\xlimin{{x\rarrow\infty \atop
    {\raise 1pt\hbox to 30pt{\rightarrowfill}}}}
\def\abs#1{\vert #1\vert}

\def\lvec#1{\rlap#1\raise 6pt\hbox{$\scriptscriptstyle\leftarrow$}}
\def\dleft{\rlap{{\it D}}\raise 8pt\hbox{$\scriptscriptstyle\Leftarrow$}}
\def\dright{\rlap{{\it
D}}\raise 8pt\hbox{$\scriptscriptstyle\Rightarrow$}}
\def\DD{\dleft\dright}
\def\lrartop#1{\rlap#1\raise 6pt\hbox{$\scriptscriptstyle
\leftrightarrow$}}
\def\omosq{\omega\_{0}\&{2}}
\def\omosqm{\omega\_{0m}\&{2}}
\def\omg{\omega\_{g}}
\def\omi{\omega\_{i}}
\def\tone{t\_{1}}
\def\ttwo{t\_{2}}
\def\eff{\vec\Im}
\def\A{\vec A}
\def\va{\vec a}

\def\vv{\vec v}
\def\vvt{\vec v(t)}
\def\vr{\vec r}
\def\vrt{\vec r(t)}
\def\dvrt{\delta\vec r(t)}
\def\vrp{\vec r\_{p}}
\def\vrc{\vec r\_{c}}
\def\aex{a\_{ex}}
\def\acu{a\_{cu}}
\def\alam{a\_{\lambda}}
\def\sk{S_k}
\def\hk{H_k}
\def\sp{S_p}
\def\aver#1{{T^{-1}}\int #1~dt}
\def\avertot#1{{T^{-1}}\int\_{\tone}\&{\ttwo} #1~dt}
\def\rder#1{\vr\&{(#1)}}
\def\lpd#1#2{\left\(\pd{\lk}{\rder{#1}}\right\)^{(#2)}}
\def\lspd#1#2{\left\(\pd{L}{\rder{#1}}\right\)^{(#2)}}

\def\lp#1{\pd{\lk}{\rder{#1}}}
\def\lsp#1{\pd{L}{\rder{#1}}}
\def\vro{\vec \rho}
\def\vin{V\_{\infty}}
\def\ao{a\_{o}}
\def\aos{\ao\&{2}}
\def\aooas{\aos/a\&{2}}
\def\h0{H_{o}}
\def\ddt#1{{d\&{#1}\over dt\&{#1}}}

\def\rar{\rightarrow}
\def\oot{{1\over 2}}

\def\S{Sec.}
\def\vsq{v\&{2}}
\def\lag{Lagrangian}
\def\eom{equation of motion}
\def\eoms{equations of motion}
\def\Om{\Omega}

\def\sk{S_k}
\def\skc{S_k^c}

\def\lk{L_k}
\def\mk{M_k}
\def\lkc{L_k^c}
\def\apj{Astrophys. J.}
\def\AJ{Astronom. J.}
\def\mnras{Mon. Not. R. astr. Soc.}
\def\AA{Astron. Astrophys.}
\def\aj{Astron. J.}
\hsize 5.6truein
\parskip 0pt
\def\oddmargin{.3in}
\def\evenmargin{.1in}
\rjustline{WIS-92/93/Nov-Ph}
\vskip 0.5truein
\centerline{\tenb DYNAMICS WITH A NON-STANDARD}
\centerline{\tenb INERTIA-ACCELERATION RELATION: AN ALTERNATIVE}
\centerline{\tenb TO DARK MATTER IN GALACTIC SYSTEMS}
\vskip 20pt
\centerline{{\twelverm Mordehai Milgrom}}
\vskip 0.3truein
\centerline{Department of Physics}
\centerline{Weizmann Institute of Science}
\centerline{76100 Rehovot, Israel}
\vskip 30pt
\baselineskip 12pt
{\bf ABSTRACT}
 We begin to investigate particle dynamics that is
governed by a nonstandard kinetic action of a special form.
We are guided by a phenomenological
 scheme--the modified dynamics (MOND)--that
imputes the mass discrepancy, observed in galactic systems, not to
the presence of dark matter, but to a departure from Newtonian dynamics
below a certain scale of accelerations, $\ao$.
The particle's \eom\ in a potential $\phi$
is derived from an action, S, of the form
$S\propto\sk\(\vec r(t),\ao\)-\int\phi~dt. $
The limit $\ao\rar 0$ corresponds to Newtonian dynamics, and there
the kinetic action
$\sk$ must take the standard form. In the opposite limit,
$\ao\rar\infty$, we require
 $\sk\rar 0$--and more specifically, for circular orbits
 $\sk\propto\ao\&{-1}$--in order to
attain the phenomenological success of MOND.
If, to boot, $\sk$ is Galilei invariant it must be
time-non-local; indeed, it is non-local in the strong sense that it
 cannot even be a limit of a sequence of local,
higher-derivative theories, with increasing order.
This is a blessing, as such theories need not suffer from the illnesses
that are endemic to higher-derivative theories.
We comment on the possibility
 that such a modified law of motion is an effective theory
resulting from the elimination of degrees of freedom pertaining to
the universe at large (the near equality $\ao\approx c\h0$ being a
trace of that connection).
 We derive a general virial relation for bounded trajectories.
Exact solutions are obtained
 for circular orbits, which pertain to rotation curves of disk galaxies:
 The orbital
speed, $v$, and radius, $r$, are related by (the rotation curve)
$\mu(a/\ao)a=\deriv{\phi}{r}$ where $a=v\&{2}/r$, and $\m(x)$ is simply
related to the value of the kinetic action, $\sk$, for a circular
trajectory.
Expressions for the angular momentum and energy of circular
orbits are derived.
Beyond the obvious matter-of-principle differences
between the modified-law-of-motion theories treated here,
 and modifications of gravity, there are important applicative
differences.
 For example, the expression for the angular momentum--an
 invariant under temporal variations of $\ao$--is
different, and the evolution of disk galaxies, expected to ensue
such variations, would be different in the two classes of theories.
The various points we make are demonstrated on
examples of theories of this class, but we
cannot yet offer a specific theory that satisfies us on all accounts,
even at the non-relativistic level that we consider here.
We also explore, in passing,
 theories that depart from the conventional Newtonian
dynamics for very low frequencies. Such theories may be written that are
linear, and hence more amenable to solution and analysis than the
MOND theories; but, they do not fare as well
in explaining the observations.

\vskip .3truein
{\bf I. INTRODUCTION: THE GALACTIC MASS DISCREPANCY
 AND THE MODIFIED DYNAMICS}
\par
It has been posited that the observed mass discrepancy, evinced by
 galactic systems, stems from the breakdown of the laws that are
 used to analyze their dynamics. This
 contrasts with the prevalent dark-matter
 doctrine that imputes the discrepancy
to the presence of large quantities of, yet unobserved, matter.
In particular, the problem has been attributed\REFS{\mia}
{M. Milgrom, \apj\ {\bf 270}, 365 (1983).}
\REFSCON{\mib}{M. Milgrom, \apj\ {\bf 270}, 371 (1983).}
\REFSCON{\mic}{M. Milgrom, \apj\ {\bf 270}, 384 (1983).}\refsend~
to a departure from Newton laws that occurs in the limit of small
 accelerations, such as are found in galactic systems:
There exists a certain acceleration parameter, $\ao$,
much above which Newton laws apply with great accuracy. Much below $\ao$
the dynamics (of gravitational systems, at least) departs drastically
from these laws, and may be epitomized by a relation of the form
$$a\&{2}/\ao=MG/r^2, \eqn{\i} $$
which gives the gravitational acceleration, $a$, of a test particle, at a
distance $r$ from a (point) mass $M$ (instead of the usual $a=MG/r^2$).
There are different theories that might spring from this seed relation.
In fact, our purpose here is to point a new direction for
searching such a theory.
\par
Arguably, the performance of this modified nonrelativistic dynamics
(MOND) has been quite successful
in reproducing the observations of galactic systems\REFS{\kent}
{S.M. Kent, \AJ\ {\bf 93}, 816 (1987).}\REFSCON{\mirc}
{M. Milgrom, \apj\ {\bf 333}, 689 (1988).}\REFSCON{\sanrev}
{R.H. Sanders, Astron. Astrophys. Rev. {\bf 2}, 1 (1990).}
\REFSCON{\bbs}{K.G. Begemen, A.H. Broeils, and R.H. Sanders,
\mnras\ {\bf 249}, 523 (1991).}
\refsend~
 (if not altogether without dispute--see e.g.
Lake\REF{\lake}{G. Lake,  \apj\ {\bf 345}, L17 (1989).}\refend,
and the rebuttal by Milgrom\REF{\milak}{M. Milgrom, \apj\ {\bf 367}, 490
(1991).}\refend).
\par
The following predictions are particularly clear-cut and well-nigh
oblivious to the details of the underlying theory.
Most of them pertain to circular motion in axisymmetric disk galaxies.
\bk
1. The speed on a circular orbit,
supported gravitationally by a central body, becomes independent of
the orbital radius for large radii\refesh{\mib}. This implies universally
flat asymptotic rotation curves of isolated disk galaxies, in accord with
high quality observations of extended rotation
 curves\refesh{\kent,\mirc,\bbs}.
\bk
2. The asymptotic rotational speed, $\vin$,
 depends only on the total mass, $M$,
of the central body via $\vin^4=MG\ao$. This
leads to a relation of the form $\vin^4\propto L$, where $L$ is any
 luminosity parameter that is proportional to the total mass. This is to
 be compared with the observed strong $\vin -L$ correlation, known as
the Tully-Fisher relation\REF{\tf}
{R.B. Tully and J.R. Fisher, \AA\ {\bf 54}, 661 (1977).
M. Aaronson and J. Mould, \apj\ {\bf 265},1 (1983).
M. Aaronson, G. Bothun, J. Mould, J. Huchra,
R.A. Schommer, and N.E. Cornell, \apj\ {\bf 302}, 536 (1986).}\refend.\bk
3. If one approximates elliptical galaxies by isothermal spheres, MOND
predicts
\REF{\miis}{M. Milgrom, \apj\ {\bf 287}, 571 (1984).}
\refend~
 a strong correlation between the velocity dispersion
 in a galaxy, $\sigma$, and its total mass $M$:
$\sigma^4\sim MG\ao$, reproducing the observed Faber-Jackson\REF{\fj}
{S.M. Faber and R.E. Jackson, \apj\ {\bf 204}, 668 (1976).}\refend~
relation between $\sigma$ and the luminosity.
\bk
4. MOND predicts\refesh{\mib} that galaxies with very low surface
 density--which correspond to low mean accelerations--will
evince a large mass discrepancy. This has been forcibly born
 out by many recent observations of low-surface-brightness disks
\REF{\lsd}
{C. Carignan, R. Sancisi, and T.S. van Albada, \aj\ {\bf 95}, 37 (1988).
C. Carignan and S.  Beaulieu, \apj\ {\bf 347}, 192 (1989).
G. Lake, R.A. Schommer , and J.H van Gorkom, \aj\ {\bf 99}, 547
(1990).
D. Puche, C. Carignan, and A. Bosma, \aj\ {\bf 100}, 1468
(1990).
M. Jobin and C. Carignan, \aj\ {\bf 100}, 648 (1990).}\refend,
and, with less confidence, for dwarf-spheroidal satellites of
our galaxy\REF{\dsp}{C. Pryor and J. Kormendy, \aj\ {\bf 100},
 127 (1990).
M. Mateo, E. Olszewski, D.L. Welch, P. Fischer, and W. Kunkel,
\aj\ {\bf 102}, 914 (1991).}\refend.
\bk
5. According to MOND, elliptical galaxies--as modeled by isothermal
spheres--cannot have a mean acceleration much exceeding $\ao$ (or
a mean surface density much exceeding the critical surface density,
$\Sigma_o\equiv\ao G\&{-1}$). This explains the observed cutoff in the
distribution of mean surface brightnesses of elliptical galaxies,
known as the Fish
 law\REFS{\fish}{R.A. Fish, \apj\ {\bf 139}, 284 (1964).}
\REFSCON{\vdk}{P.C. van der Kruit, \AA\ {\bf 173}, 59 (1987).}\refsend~
\bk
6. It is known that bare, thin disks, supported by rotation,
are unstable to bar formation, and eventual breakup. MOND endows
such disks with added stability that enables them to survive, provided
they fall in the MOND regime, i.e. their mean acceleration is at
$\ao$ or below\REF{\stab}{M. Milgrom, \apj\ {\bf 338}, 121 (1989).
D.M. Christodoulou, \apj\ {\bf 372}, 471 (1991).}\refend.
This can explain the observed marked paucity of galactic disks
with mean surface density larger than the above $\Sigma_o$,
known as the Freeman law (see ref. \refesh{\vdk})
for a discussion of this issue).
\bk
7. Above all, detailed analysis of rotation curves, for galaxies with
high-quality data, show a very close match of the predictions of MOND
with the observed rotation curves of these
 galaxies\refesh{\kent,\mirc,\bbs}.
\par
Generally, the dynamics of systems of galaxies--such as small groups,
galaxy clusters, and the Virgo infall--which are not as well understood,
can also be explained by MOND, with no need to invoke dark
 matter\refesh{\mic}.
\par
In the framework of MOND, all these tell us that a modification is called
for when dealing with bound orbits.
 There is no observational information
on unbound (scattering) trajectories, for massive particles. However,
we may gain insight from
observations of gravitational lensing of light from distant galaxies,
by foreground clusters of galaxies\REF{\lens}{See e.g. J.A. Tyson,
F. Valdes, and R.A. Wenk, Astrophys. J. Lett. {\bf 349}, L1 (1990).
J.A. Tyson, Phys. Today, June, 24 (1992).}\refend.
 These however concern relativistic trajectories, which require
 a relativistic extension of MOND for a proper treatment
(see below and \S IV).
Henceforth we assume MOND for all
types of nonrelativistic trajectories.
\par
The value of $\ao$ has been determined, with concurrent results, by
several independent methods; these
rest on the twofold role of $\ao$
 as the borderline acceleration (based
on predictions 5 and 6 above), and as the scale of acceleration in the
deep MOND limit (based on predictions 2 and 3). The best determination
comes from the detailed studies of rotation
 curves\refesh{\kent,\mirc,\bbs}, and is
$$\ao\sim(1-1.5)10\&{-8}~cm~s\&{-2},  \eqn{\anought} $$
assuming the value of $75 \abunit$ for the Hubble constant.
Values of $a/\ao$ as small as 0.1 have been probed in studies of
rotation curves; for the dynamics of the infall into the Virgo cluster
one has typically  $a/\ao\approx 0.01$.

\par
MOND, which at the moment, is but a phenomenological theory,
may be interpreted as either a modification of inertia, or as a
modification of gravity that leaves the law of motion
 intact\refesh{\mia}.
True, General Relativity has taught us that gravity and inertia
 are not quite separable
on the fundamental level. For instance,
 the Einstein field equations for
the gravitational field imply the geodesic law of motion.
 However, at the level at which we work--i.e. a
 non-relativistic effective theory--it
 does make sense to distinguish between the interpretations.
By the former we mean, foremost, a modification the must be implemented
for whatever combination of forces determines the motion.
\par
In default of a full-fledged theory
Milgrom\refesh{\mia,\mib,\mic}
 had initially used a working extrapolating
\eom\ of the form
$$\m(a/\ao)\va=\va\_{N}\equiv -\grad\phi,  \eqn{\form}$$
to calculate the acceleration $\va$ from the conventional potential
$\phi$ ($\va\_{N}$ is the Newtonian acceleration).
 The extrapolating function, $\m(x)$, satisfies
$\m(x)\xlimin 1$, so that Newtonian dynamics
is attained in the limit of large accelerations. In the deep MOND limit
we have $\m(x)\propto x$, for $x\ll 1$ (and $\ao$ is normalized such that
$\m(x)\approx x$),
 so that eq.\i results, and all
the above-mentioned predictions follow.
Relation \form\ is noncommittal on the matter of interpretation
 (see ref. \refesh{\mia}).
 As it is written it may be viewed as a modification
of the law of motion for arbitrary potential $\phi$, or, inverted to read
$\va=-\n(\abs{\grad\phi}/\ao)\grad\phi$, and assumed to apply only
for gravitational potential, it may be viewed, instead,
 as a modification of
gravity.
\par
Bekenstein and Milgrom\REF{\bm}
{J.D. Bekenstein and M. Milgrom,  \apj\ {\bf 286}, 7 (1984).}\refend~
have then devised a nonrelativistic
 formulation of MOND that is expressly a modification of gravity:
It replaces the Poisson equation ($\div\grad\phi=4\pi G\varrho$) by
$$\div\(\m(\vert\grad\phi\vert/\ao)\grad\phi\)=4\pi G\varrho,
  \eqn{\mbm}$$
which determines the gravitational potential, $\phi$, generated
by the mass-density
distribution $\varrho$.
 The acceleration of a test particle is then given by
$-\grad\phi$.
 This formulation is derivable from an action
 principle, enjoys the usual conservation laws,
and has been shown to be satisfactory on other important
 accounts\refesh{\bm}. It reduces to eq.\form\
in cases of one-dimensional
 symmetry, and, in fact, even for less symmetric
systems such as disk galaxies, has been shown to predict very similar
 rotation curves as eq.\form .
 Notwithstanding the superior
standing of eq.\mbm, it is
eq.\form that has been used in all
the rotation-curve tests of
 MOND\refesh{\kent,\mirc,\bbs}--being
much wieldier.
\par
There are several relativistic theories with MOND as their
 non-relativistic limit\refesh{\bm}\REF
{\biba}{J.D. Bekenstein,
in {\it General Relativity and Astrophysics},
 eds. A. Coley, C. Dyer, and T. Tupper, ( World
Scientific, Singapore, p. 64) (1988).
J.D. Bekenstein,  Phys. Lett. B, {\bf 202}, 497 (1988).
R.H. Sanders, \mnras\ {\bf 223}, 539 (1986).
R.H. Sanders, \mnras\ {\bf 235}, 105 (1988).
J.D. Bekenstein, Sixth Marcel Grossman Conference on General Relativity,
Kyoto (1991).}\refend; none is without problems. They are all based
on modification of gravity, in that they can be formulated as a
modification of the Einstein field equations for the metric, with
particles moving on geodesics. Obviously, this entails an indirect
effect on other forces when gravity is important; but, inasmuch
as gravity can be neglected, they do not entail a modification of the
motion of particles subjected to other forces.
\par
In this paper we begin to discuss formulations of MOND that,
in contrast, may be viewd as a modifications of
 the law of motion of a particle in a given potential field.
The main implication here is that particle dynamics is affected for
 whichever combination of forces is at play.
\par
We derive the dynamics from an action principle, with an
 action
$S=\sp+\sk$,
where the potential part, $\sp$, takes the conventional form,
while the kinetic action, $\sk$, carries the modification.
The latter
 is functional of the trajectory of the particle, and is constructed
using the acceleration constant $\ao$ as the only extra parameter.
The theory must approach the Newtonian dynamics in the limit
$\ao\rar 0$. When $\ao$ is not much smaller then
all the quantities of the dimensions of acceleration,
MOND takes
 its effect. The deep MOND limit corresponds to $\ao\rar\infty$.
To obtain the behavior epitomized by eq.\i --or, to have asymptotic
flat rotation curves for isolated galaxies--we expect
$\sk\propto \ao\&{-1}$ in this limit,
 at least for circular trajectories.
\par
Thus--in departure from the pristine statement of the MOND
 hypothesis--deviations from Newtonian dynamics appear, here,
 not only when the
momentary acceleration on the trajectory is much smaller than $\ao$.
The special role of the acceleration is highlighted only through the
introduction of the constant $\ao$ of these dimensions.
On a circular, constant-speed trajectories, all the quantities
with the dimensions of acceleration that can be built
equal the acceleration itself. For such trajectories
the acceleration only has
 relevancy for MOND (as is the case for
 rotation curves of disk galaxies).
\par
The main purposes of this paper are:
(i) To adumbrate a possible rationale for MOND, which we do in \S\ II.
(ii) To demonstrate
 that it is possible to formulate MOND as a modification
of the law of motion--in contrast to a modification of gravity--and
to expound some general properties of theories based on such a
 modification (\S\ III).
(iii) To point out significant differences
between the two interpretations, and to highlight the observational
implications of these differences (\S\ VI).
(iv) To present general, exact results pertaining to circular
 motion, and point their relevance to the dynamics of disk galaxies.
In particular, we give a general expression for the rotation curve
of an axisymmetric disk galaxy, and derive simple expressions for the
energy and angular momentum of circular orbits in \lag\
theories; this we do in \S\ IV.
\par
Some of the points we want to explicate
 become more transparent in the
context of a theory where the dynamics is modified in the limit of
low frequencies, instead of the limit of small accelerations.
Theories of this class may be constructed that are linear, and thus
 more amenable
to analysis and solution; alas, as potential alternatives to dark matter
in galactic systems, they seem to be at some odds with
the observations. We discuss these, in some detail,
in \S\ V, both for their own possible merit, and as heuristic
auxiliaries.
Finally, we list some of the desiderata that we have not been able
to implement yet,
 and some questions that remain open
 (\S\ VI).
\par
We cannot pinpoint a specific theory. There is a large
variety of theories that, as far as we have checked, are consistent with
the available, relevant data (especially RCs). Also, we can offer no
single model that possesses all the features we would ultimately require
from a theory (see \S\ VI).
\par
This approach opens
 a new course for searching for a sorely needed, relativistic
extension of MOND.
\vskip .3truein
{\bf II. PREAMBLE: THE POSSIBLE PROVENANCE OF THE MODIFIED
 DYNAMICS}
\par
Various considerations lead us to suspect that MOND is an effective
approximation of a theory at a deeper stratum.
Here, a most germane observation\refesh{\mia} is that the deduced
value of $\ao$ \(eq.\anought\) is of the same order as $c\h0$,
where $\h0$ is the present expansion rate of the universe (the
Hubble constant), and $c$ the speed of light.
(For $\h0=75~\abunit,~ c\h0\sim 7\times10\&{-8}cm~s\&{-2}$.)
This near equality may betoken an effect of the universe as a whole on
local dynamics, on systems that are small on the cosmological scale.
\par
There are, in fact, several quantities with the dimensions of an
acceleration, that can be constructed from cosmological parameters\REF
{\micomm}{M. Milgrom, Comm. Astrophys. {\bf 13}, 215 (1989).}\refend,
beside the above expansion parameter
$$\aex\equiv c\h0.    \eqn{\aexp}$$
Another is
$$\acu\equiv c^2/R_c,   \eqn{\rcur}$$
where $R_c$ is the radius of spatial curvature of the universe;
yet another is
$$\alam\equiv c^2\vert\lambda\vert\&{1/2},  \eqn{\alambda}$$
with $\lambda$ the cosmological constant
($\lambda\equiv\Lambda/3c\&{2}$, where $\Lambda$ is sometimes used).
\par
Only $\aex$ is observationally determined with some accuracy, as the
Hubble constant has been measured to lie between about 50 and 100
$\abunit$ (see Huchra\REF{\hubble}{J.P. Huchra, Science
{\bf 256}, 321 (1992).}\refend~ for a review).
On $R_c$ we have only a lower limit of the order of
$c/\h0$, and on $\vert\lambda\vert$ we have only an upper limit
\REFS{\coscon}{S. Weinberg, Rev. Mod. Phys. {\bf 61}, 1 (1989).}\REFSCON
{\cntwo}{S.M. Carroll, W.H. Press, and E.L. Turner, Ann. Rev. Astron.
Astrophys. {\bf 30}, 499 (1992).}\refsend~
 of order $(\h0/c)\&{2}$.
So, only upper limits exist on $\acu$ and $\alam$, and these are of
the order of $\aex$.
In a Friedmannian universe, with vanishing cosmological constant,
one has
$$\acu/\aex=\vert 1-\Omega\vert\&{1/2},  \eqn{\rat} $$
where $\Omega$ is the ratio of the mean density in the universe to the
closure density. (We use these relations, derived from conventional
cosmology, only heuristically, as MOND may entail a modification
of the cosmological equations.)
So, if today $\Omega$ is not very near 1, $\acu$ and $\aex$
are comparable.
If the mean mass density falls much shorter of closure--as would
be in keeping with the basic premise of MOND--but a non-zero
cosmological constant renders the universe flat--as
 would conform with inflationary models\REF{\lala}
{See e.g.
P.J.E. Peebles, \apj\ {\bf 284}, 439 (1984); G. Efstathiou, W.J.
Sutherland, and S.J. Maddox, Nature {\bf 348}, 705 (1990).}\refend~
(see also ref. \refesh{\cntwo})
--we have today
$$\alam\approx\aex.  \eqn{\appp}$$
\par
We see then that the near equality $\ao\sim \aex$ may be fortuitous,
with $\ao$ being really a proxy for $\acu$, or for $\alam$
 (or for another cosmic parameter).
The identification of the parameter behind $\ao$ would be an
important step toward constructing an underlying theory for MOND;
it is also of great impact in connection with formation of
structure in the universe, and the subsequent evolution of galaxies,
and galactic systems\refesh{\micomm}: Whereas $\alam$ is a veritable
constant, $\aex$, and $\acu$ vary with cosmic time (in different
ways). The matter has also antropic bearings.
\par
The Newtonian limit, which
corresponds to $\ao=0$, would be realized, according to the above
identifications of $\ao$, when the universe is static
($\aex=0$), flat ($\acu=0$), or characterized by
 a vanishing cosmological constant
($\alam=0$), respectively.
\par
It is clear that
to some degree the various cosmological parameters
must appear in local dynamics.
A finite curvature radius of the universe would enter, at least
through the boundary conditions for the Einstein field equations,
and will modify the local gravitational field. Likewise, a non-zero
cosmological constant modifies the distance dependence of Newtonian
 gravity. The expansion of the universe also enters, for example
by slowing down otherwise-free massive particles (or red-shifting
free photons). All these are not, however, what we need for MOND.
These effects are only appreciable over distances or times
of cosmological scales--not relevant for non-relativistic
 galactic systems and phenomena.
We thus do not expect $\h0$, $R_c$, and $\lambda$ to enter local dynamics
as a time or length scale.
 On the other
hand, $\ao$ is of the the same order as the galactic parameter
of the same dimensions. So, if $\ao$ does remain as a
vestige  of the underlying mechanism, it will be the only such parameter
 important for non-relativistic galactic phenomena.
In contrast, {\it relativistic} phenomena
characterized by accelerations of order $\ao$ or less, also
involve scale length of cosmological magnitude.
For example, a photon trajectory passing a distance $r$ from a mass $M$,
such that $MG/r\&{2}\leq\ao$, has a curvature radius of cosmological
scale (see more in \S IV). So, the fact that only $\ao$ appears in MOND
may be peculiar to the non-relativistic nature of this theory.
\par
 The following analogy may help illuminate the
situation we have in mind: Had all our knowledge come from experiments
in a small, closed
laboratory on the earth's surface, our dynamics would have
involved a
``universal'' constant $\vec g$--the free-fall acceleration.
Unaware of the action of the earth, we would find a relation
$\vec F=m(\va- \vec g)$ between the applied force $\vec F$, and
the acceleration $\va$ of a particle.
Actually, $\vec g$ encapsules the effects of the mass,
$M\_{\oplus}$, and radius, $R\_{\oplus}$, of the earth. $R\_{\oplus}$,
by itself, appears conspicuously in the dynamics
  only on scales comparable
with itself; $\vec g$ alone enters the results of
very-small-scale experiments characterized
by accelerations not much larger then $g$.
 (The escape speed here plays the part
of the speed of light in connection with MOND:
 $g=v\&{2}\_{es}/2R\_{\oplus}$.)
The effective departure here
 is of a gravitational origin, but we may
describe it as a modification of inertia because
 it is in effect for whatever
combination of forces $\vec F$ represents, even
when there are no ``local'' sources of gravity (i.e. sources
in addition to the earth). We would thus have no reason to consider
it a modification of gravity.
 It may well be that the departure inherent in
MOND is also of a gravitational origin (and will then automatically
preserve the weak equivalence principle in the effective theory).
\par
But why, at all,
should $\h0$, $R_c$, or $\lambda$ enter local dynamics as
an acceleration parameter?
We can only offer a vague possible rationale:
Even within the purview of conventional physics we expect
a difference of sorts in the dynamics of particles
with accelerations above and below $\ao$, if $\ao$ stands for
one of the above cosmological parameters.
Consider, for example, a electric charge accelerated with acceleration
 $a$; the lower radius of its
radiation zone is $r\_{rad}=c^2/a$. To say that $a\ll\acu$, for example,
 is to say
that $r\_{rad}$ is much larger than the radius of curvature of
universe.
 In a closed universe, for example, this would mean that there in not
even room for a radiation zone. Clearly, the radiation pattern
must differ from the case where $a\gg\acu$ for which
 the curvature is hardly
felt within the near field.
More generally, an accelerated particle carries a causal structure
that includes an event horizon\REF{\sdhji}
{See, e.g. W. Misner, K.S. Thorne, and J.A. Wheeler
{\it Gravitation}, W.H. Freeman, N.Y. (1973).}\refend,
 whose scale is
$\ell=c^2/a$: This is the distance to the horizon; this is the space-time
distance within which a freely falling frame can be erected; this is
the typical wavelength of the Unruh black-body radiation, etc..
For accelerations that are low in the context of MOND ($a\ll\ao$),
the quality of the universe that is behind $\ao$ is felt
within $\ell$. Conversly, we surmise that $\ell$ is imprinted on the
conjectured inertia field (e.g. by modifying the spectrum of the
vacuum fields) in a way that is then sensitive to the acceleration
of a test particle interacting with this field.
\par
 All the above considerations
 may also usher MOND in as an effective theory of gravity.
\par
We thus envisage inertia as resulting from the interaction of the
accelerated body with some agent field, perhaps having to do with the
vacuum fields, perhaps with an ``inertia field'' whose source is
 matter in the ``rest of the universe''--in the spirit
of Mach's principle.
 If this field bears the imprint
of at least one of the cosmic parameters, an effective description
of dynamics in which this inertia field is eliminated, may result
in a scheme like MOND.
\par
We have not been able to put these vague ideas into a concrete use,
and will have to make do, in the rest of this paper, with the study
of effective theories.
Nonetheless, we keep this conjectured origin
 before our eyes as a guide of sorts
on what MOND may entail, beside the implications for the structure
of galactic systems, which have been the phenomenological anchor
for MOND.
For example, such considerations would lead us reckon with the effects
of an $\ao$ that varies with cosmic time, as alluded to above.
We may also suspect that if MOND, as we now formulate it,
is only an effective theory delimited by cosmological factors, then
it cannot be used to describe cosmology itself (as we would not
describe the motion of earth satellites with a constant-$g$ dynamics
in the above analogy).
More generally, we noted above that
 MOND as described above--with $\ao$ as the only
relevant parameter--may breakdown for any relativistic phenomenon
with acceleration below $\ao$ (more in \S VI).
\par
There are attributes of the  conventional laws of motion that we cannot
take for granted in light of what we said above on the
 possible origin of the modified dynamics.
For example, it is not obvious that the effective theory describing
MOND is derivable from an action principle.
The question of which symmetries (conservation laws) the theory
enjoys also becomes open.
We shall certainly want to preserve the six cherished symmetries
of space; to wit, the translations and rotations (or more
generally the six isometries of a maximally symmetric
3-D space); what we said above raises no question with regard to these:
the underpinnings of the effective theory are not expected to define a
preferred position or direction.
The same is true of parity.
Regarding time reversal symmetry,
if the expansion of the universe is in the heart
of MOND, the fact that the universe has a well defined time arrow
might well result in a strong TR asymmetry in the MOND regime,
 just as physics in an external magnetic field is TR
 non-symmetric.
It would be very difficult to ascertain a breakdown of TR symmetry in
the province of the galaxies. Many potential manifestations of this are
forbidden by other symmetries, and there are other balking factors.
The same is true of the time-translation symmetry. Here
our guess would be that time translation might be affected through the
introduction of the scale of time ($\h0\&{-1}$), and this is not
 interesting, in the context of MOND,
as it would lead to deviations only on very long time scales.
\par
Similarly, we are at sea regarding full Galilei invariance.
While the ultimate underlying theory may well be Galilei
(Poincare) invariant, or enjoy even a higher symmetry,
 this is not necessarily the case for the effective theory that
 results from the elimination of some degrees of freedom:
the surmised ``inertia field'', which comprises the hidden degrees of
freedom, defines a preferred frame.
Perhaps the effective theory does enjoy some generalization
of Galilei invariance  that involves $\ao$
as a parameter, and that reduces to Galilei invariance in the limit
$\ao=0$.
Imagine, for example, an accelerated observer in a curved space.
As long as its acceleration is much larger than $\acu$ of eq.\rcur,
so that its Unruh wavelength, or the size of the region affecting the
dynamics, is much smaller than the radius of
 curvature, he will not detect departure from Lorentz invariance
due to the curvature; for $a\ll\acu$ the non-zero curvature distorts
the Unruh spectrum.
\par
As phenomenology does not yet require that we relinquish any of the above
 symmetries,
and as we do not have any concrete theoretical impetus to do so,
 we shall retain them in the following discussion.
\par
We shall concentrate on
effective theories of motion that are derived from an action principle,
and conform to the MOND hypotheses.
Such effective theories, which are obtained by eliminating some
degrees of freedom from a more fundamental theory, are, in many
cases, non-local; i.e., they are not derivable from an action that
is an integral over a simple \lag\ that is a function of only
a finite number of time derivatives of the trajectory.
A well known case in point is the Wheeler-Feynman formulation of
electrodynamics\REF{\wf}{J.A. Wheeler and R.P. Feynman, Rev. Mod. Phys.
{\bf 21}, 425 (1949).}\refend, which results from the elimination
of the electromagnetic field, and is non-local when only the particle
coordinates remain as degrees of freedom.
We shall show--without bringing to bear the origin of MOND-- that
 an acceptable theory of MOND must, in fact, be strongly
non-local.

\vskip .3truein
{\bf III. GENERAL MOND ACTIONS}
\par
One starts by assuming that the motion of a non-relativistic
 test particle, in a static
potential field $\phi$, is governed by an \eom\ of the form
$$m\A=-\grad\phi,  \eqn{\vii} $$
where $m$ is the mass of the particle, and
 $\A$ (replacing the acceleration
 $\va$ in Newtonian dynamics) depends only
on the trajectory $\vrt$ (not on the potential); its value is, in
general, a functional of the whole trajectory, and a function of
the momentary position on the trajectory, parameterized by the time t.
Of $\A$ we further require the following:
(i) That it be independent of $m$, so as to
retain the weak equivalence principle. In what follows we take a unit
mass for the particle.
(ii) According to the MOND assumption,
$\A$ may be constructed with the aid of the acceleration constant $\ao$,
as the only dimensional constant.
This is assumed to be the case in the non-relativistic limit.
(iii) $\A$ has to approach the Newtonian expression, $\va$, in the limit
$\ao\rar 0$, for all trajectories,
 to achieve correspondence with Newtonian dynamics.
(iv) The deep MOND limit corresponds to $\ao\rar\infty$; phenomenology
points to $\A\propto\ao\&{-1}$, in this limit.
We do not know that this shoul hold universally (i.e. for all sort
of trajectories etc.), we shall assume this behavior for circular orbits,
and for general trajectories assume only $\A\rar 0$.
We also do not have observational information on the very limit
$\ao\rar\infty$;
reliable information exists only down to $a/\ao\sim 0.1$. Here we shall
assume this everywhere for large $\ao$, but most of what we say
is independent of this assumption.
(v) We assume $\ao$ to be time independent, and so we are free to
assume that
$\A$ does not depend explicitly on time.
If $\ao$ actually varies with cosmic time, our results may be considered
the lowest order in an adiabatic approximation.
Explicit time dependence could also enter $\A$ if the particle has
variable mass, which is simple to include (but which we bar).
(vi) $\A$ is invariant under translations, and transforms as a vector
under rotations of the trajectory.
(vii) We do insist on full Galilei invariance.
\par
Indication that we need require some invariance connected
with a velocity boost could have come from the study of unbound
trajectories.
 When a fast, unbound particle of velocity $v$
scatters gravitationally off some mass, with pericenter distance
$r$, the characteristic acceleration around pericenter is
 $a\approx v\_{es}\&{2}/r$, where $v\_{es}$ is the escape speed from $r$.
We can, however, with the aid of $v$,
 construct other quantities with the dimensions of acceleration, e.g.
$a\_{n}=(v\&{n}\ddt{n}a)\&{1/(n+1)}$. Around pericenter,
$a\_{n}\approx (v/v\_{es})\&{2n/(n+1)}a$, which
may be much larger than $a$ itself.
This might lead, in theories such as we discuss here, to
much less of a mass discrepancy for such unbound
particles, compared with that manifested by
 bound particles; for the latter, all quantities
of the dimensions of acceleration are of order $a$.
It is the appearance of the velocity $v$ that engenders such terms.
Galilei invariance would bar the appearance of $v$ in the
\eom, and all the permissible quantities with dimensions of
acceleration are of the order of the acceleration, near pericenter.
We may then expect that bound, and scattering test particles will
point to mass discrepancies of the same order ($\ao/a$).
Observations of lensing by galaxy clusters may illuminate this question
(for these  $v/v\_{es}$ is typically a few hundred), pending the
understanding of the relativistic extension of MOND.
\par
While, for convenience, we treat motion in a static potential, all
our results apply to the relative motion in the center-of-mass frame,
of a system of two particles , interacting
through a potential that depends on the relative position. So Galilei
invariance of $\A$ for the particles assures invariance for the system.
\par
Equation \form\ satisfies all the above requirements, but does not
conform to another the we now add; viz., that the \eom\ be derivable
from an action principle. As far as we can see, there is nothing amiss,
in principle, with effective
 \eoms\ that are not so derivable. If their symmetries
do not lead to conservation laws, this may just reflect interaction
with the hidden degrees of freedom. In fact, exactly such effective
 theories have been proposed as models for dissipative systems\REF{\leca}
{See e.g. A.O. Caldeira and A.J. Legget, Ann. Phys. {\bf 149}, 347 (1983)
and references therein.}\refend.
That being as it may, our purpose here is to investigate \eoms\ with an
action underpinning. This is a strong constraint that enables
us to say much about the theory.
One result of this assumption is that a theory whose kinetic
 action is Galilei invariant, and has the correct Newtonian limit,
and the required MOND limit, cannot be local; i.e. $\A$ in eq.\vii \
cannot be a function of a finite number of time derivatives of $\vrt$.
(From this we understand why eq.\form , which is local, cannot be
derived from an action.) Other results concern a virial relation,
 conservation
laws, and adiabatic invariants. In yet another we shall show that
circular orbits are necessarily solutions of the \eom\ in a potential
with the appropriate symmetry, and derive explicit expressions
for the rotational speed.
\par
We formulate the action principle as follows: There exists a kinetic
action functional $\sk\(\vrt,\tone,\ttwo,\ao\)$
(not necessarily \lag ) that is a functional of
the trajectory, $\vrt$, given between two end times $\tone$, and
$\ttwo$. Further assume that
the variation of $\sk$ under
a change $\dvrt$ may be written as
$$\d\sk=-\avertot{\A(t)\cdot\dvrt}+
{} ~T^{-1}\(\eff\cdot\dvrt\)\&{\ttwo}\_{\tone}, \eqn{\vi} $$
where $\A$ does not depend on the end points.
The second term on the
 right-hand side is (up to the $T$ normalization)
 the difference between the values of
$\eff\cdot\dvrt$, at the two end times;
 $\eff$ is some linear vector operator,
acting on functions of $t$, \(and may depend on $\vrt$\).
We then consider only \eoms\vii whose kinetic part $\A$ can be obtained
in this way from an action functional.
The normalization by the time interval $T\equiv \ttwo-\tone$
is introduced for convenience, so as to render the action
finite when $t\rar\infty$; it does not affect the \eom.
\par
All the above requirements we made of $\A$ are now carried, {\it mutatis
mutandis}, to the kinetic action $\sk$.
\par
Because we want the kinetic part of the \eom\ to bear the modification
that MOND entails, we choose to discuss the kinetic part separately.
The full action of the theory is
$$S=\sk+\sp,     \eqn{\ii} $$
where the potential part has the usual form (except for our
 normalization):
$$ \sp=-\avertot{\phi}.  \eqn{\iii} $$
The variation of the full action is
$$\d S=-\avertot{\(\A(t)+\grad\phi\)\cdot\dvrt}+
{} ~T^{-1}\(\eff\cdot\dvrt\)\&{\ttwo}\_{\tone}. \eqn{\vis} $$
\par
Note that we do not define the physical trajectories as extrema of the
action; that would be the case only if one restricts himself to
 variation, $\dvrt$, for which the ends term vanishes, which is usually
done by fiat. The trajectory may well be an extremum, but we need not
 and wish not make such requirements.
\par
Standard Lagrangian theories are special instances (see below), but the
above definition will suit us in more general cases.
When the theory is non-local, it usually makes sense
only when we take $\tone$ to $-\infty$, and $\ttwo$ to $+\infty$.
In this case the \eom\ is a non-local equation for the full world-line
of the particle.
A finite trajectory (specified between finite times) is permitted
if it is a segment of a world-line that solves the \eom .
An important question regarding a given theory is how many (vector)
parameters are needed to specify a permitted
world line; this is also the number of initial conditions that one has to
specify for a finite time $t$.
\par
Consider now variations of $S$ under coordinate transformations.
For solutions of the \eom , only the ends term remains in eq.\vis,
 and it is thus
clear that $\eff$ generates conserved quantities, when underlying
symmetries of $S$ are present. For example, if $S$ is invariant under
translations $\vr\rar\vr+\d \vec\epsilon$, we see that the difference
of the values of $\eff(1)$ ($\eff$ acting on 1)
between the two end points
 must vanish. This quantity may be identified as the momentum.
Similarly, if $S$ is invariant under rotations about a direction
$\vec n$, since $\d\vr=\epsilon\vec n\times\vr$, we find that
$-\vec n\cdot\(\eff\times\vrt\)$ is the same at the two end points.
Thus we identify $-\eff\times\vrt$ with the angular momentum.
\par
Now take a dilatational increment in a solution of the \eom:
$\dvrt=\epsilon\vrt$. For trajectories on which $\eff\cdot\vrt$
does not increase as fast as $t$, at $t\rar\pm\infty$
we get, from eq.\vi,
  for the variation of $\sk$ (for the full world line
$T\rar\infty$)
$$\d\sk=-\epsilon \langle\A\cdot\vrp\rangle, \eqn{\wxii}$$
where $\langle\rangle$ is the time average over the trajectory.
This certainly holds
for periodic trajectories, but surmisedly for any bounded
trajectory.
We now calculate $\d\sk$ differently:
On dimensional grounds
$$\sk\(\l\vrt,\l\ao\)=\l\&{2}\sk\(\vrt,\ao\), \eqn{\wpnmb} $$
for a constant $\l$.
Putting $\l=1+\epsilon$,
 and identifying the first order terms in
$\epsilon$, we get, as in Euler's theorem for homogeneous functions,
$$\d\sk+\epsilon\ao\pd{\sk}{\ao}=2\epsilon\sk.  \eqn{\xiv}$$
Comparing with expression\wxii for $\d\sk$
we get
$$-\langle\A\cdot\vr\rangle=2\sk-\ao\pd{\sk}{\ao},\eqn{\wxv}$$
and from the equation of motion
$$\sk(1+\oot\hat\sk)=\oot\langle\vr\cdot\grad\phi\rangle, \eqn{\wxvi}$$
where $\hat\sk\equiv-\pd{ln~\sk}{ln~\ao}$.
This is a generalization of the virial relation (in the standard
Newtonian case $\sk$ is the mean kinetic
energy, with our normalization).
\par
More generally, if $\sk$ depends on a number of
 dimensional constants, $C\_{\ell},~~\ell=1,...,I$,
with dimensions $(length)\&{\a\_{\ell}}\times (time)\&{\b\_{\ell}}$,
respectively,
then $\hat\sk$ in eq.\wxvi is replaced by
$-\sum\_{\ell=1}\&{I}\a\_{\ell}\pd{ln~\sk}{ln~C\_{\ell}}$.
(If we replace $\sk$ with $S$, and put
$\phi=0$ we get a more general relation that reduces to the former
when $S=\sk+\phi$ with $\phi$ depending only on $\vr$, because then
we must have $\phi=C^2f(\vr/r\_{0})$ with $C$ and $r\_{0}$ of dimensions
of velocity and length, respectively.)
\par
In Newtonian dynamics $\A=\va$ is linear in $\vrt$.
In a MOND theory this may not be: From dimensional arguments
$$\A\(\l\vrt,\l\ao\)=\l\A\(\vrt,\ao\),  \eqn{\pnmf}$$
or, taking $\l=\ao\&{-1}$,
$$\A\(\vrt,\ao\)=\ao\A\(\ao\&{-1}\vrt,1\).  \eqn{\abcd} $$
If $\A$ is linear in $\vrt$, it is
independent of $\ao$ altogether.
\par
It may be more convenient, sometimes, to write the \eom\
in terms of the expansion coefficients of $\vrt$
in some basis of complete functions, instead of $\vrt$ (which is
itself
the expansion coefficient of the trajectory in the functions $\d(t'-t)$).
This is not a common practice because (i) for time-local, \lag\
theories the locality is manifest only when $t$ is used as an independent
 variable
(ii) $t$ is a natural choice when the potential is time independent.
If we forgo locality anyway,
 the benefits of using a different decomposition of
the trajectory may outweigh the drawbacks.
For example, we may use the Fourier transform, $\vro(\omega)$, of
$\vrt$. As a functional of $\vro$ the kinetic  action may take a
particularly simple form. We discuss examples below.
\par
It is well known that theories with higher derivatives tend to suffer
 from various diseases. we discuss these and their prevention
at the end of this section, after we have considered a few examples.
\vskip .15truein
\bk
{\it LAGRANGIAN THEORIES}
\par
We now specialize to kinetic actions that are derivable from a \lag\
function:
$$\sk=T^{-1}\int\_{\tone}\&{\ttwo}\lk~dt \eqn{\x} $$
(the total \lag\ being $L=\lk-\phi$).
We assume $\phi$ to depend only on $\vr$ (and, in particular, not to
depend explicitly on the time).
In this case $\A$ is given by  the standard Euler-Lagrange expression
$$\A=\sum\&{N}\_{i=1}(-1)\&{i-1}\lpd{i}{i},  \eqn{\rona} $$
and the operator $\eff$ is
$$\eff=\sum\&{N}\_{m=1}\vec p\_{m}D^{m-1},  \eqn{\ronb} $$
with
$$\vec p\_{m}\equiv
 \sum\_{i=m}\&{N}(-1)\&{i-m}\lpd{i}{i-m}.    \eqn{\xxvii} $$
Here, $D\equiv\ddt{}$, and
 a superscript $(i)$ stands for the $i$th time derivative.
\par
If the number, $N$, of derivatives on which $\lk$ depends is finite, the
theory is local. Even when it is non-local, the theory may be a limit
of a sequence of local ones
 with \lag s $\lk\&{N}$, $N\rar\infty$.
(A common example is an $\lk$ that can be
expanded in a power series in some parameter--$\ao$ in our case--where
the coefficients are function of a finite number, $N$, of derivatives.)
 When it
is not even that we call the theory strongly non-local.
For such theories
 the sums in the expressions
 above do not converge for at least some
 trajectories. These expressions may still be of value, as they may
 converge for some trajectories, and may be analytically continued for
even more trajectories.
 A different expression for
$\A$ must then be given, to establish an \eom .
\par
As explained above, translation invariance implies the conservation
of the momentum $\eff(1)$, which here is just $\vec p\_{1}$ (this can
 also be seen from the fact that $\A=\deriv{\vec p\_{1}}{t}$).
(In a many-body system with bodies of masses $m\_{i}$, and potentials
that depend on coordinate differences, $\sum\_{i}m\_{i}\A\_{i}=0$.)
Similarly, the angular momentum--whose component along
 $\vec n$ is conserved
when $\phi$ is symmetric under rotations about $\vec n$--is given
by
$$\vec J=-\eff\times\vrt=\sum\_{m=1}\&{N}\rder{m-1}\times\vec p\_{m}
=\sum\_{m=1}\&{N}\rder{m-1}\times
\sum\_{i=m}\&{N}(-1)\&{i-m}\lpd{i}{i-m}, \eqn{\xxv} $$
or
$$\vec J=\sum\_{i=1}\&{N}\sum\_{m=1}\&{i}(-1)\&{i-m}\rder{m-1}\times
\lpd{i}{i-m}. \eqn{\xxva}  $$
This also follows straightforwardly from the fact that
 $\vr\times\A=\deriv{\vec J}{t}$, so the rotational symmetry of $\phi$
\(which reads $\grad\phi\cdot(\vec n\times\vr)=0$\)
 implies, from the \eom,
$\deriv{(\vec n\cdot\vec J)}{t}=0$.
\par
As in standard dynamics,
if the potential is rotationally symmetric about $\vec n$,
 $\vec n\cdot\vec J$ is conserved,
even when the \lag\ depends explicitly on time, such as when
$\ao$ varies with cosmological time. Obviously, this has implications
for galaxy evolution that is driven by such variations.
\par
If $L$ does not depend explicitly on time,
 the energy $H$ is conserved, where
$$H=\sum\_{i=1}\&{N}\sum\_{m=1}\&{i}
(-1)\&{m-1}\rder{i-m+1}\cdot \lspd{i}{m-1}
{} ~-L.  \eqn{\xxiv}$$
We separate $H=\hk+\phi$, and call $\hk$ the kinetic energy.
The conservation of $H$ can be deduced  by noting that
$\vv\cdot\A=\dot\hk+\pd{\lk}{t}$, so that, from the \eom , if
$\pd{\lk}{t}=0$, we have $\dot\hk+\dot\phi=0$.
\par
If we add to a total derivative to $L$, then
 $\A,~H,~,p\_{1}$, and $\vec J$
remain the same, but $p\_{i}$ for $i>1$, and hence $\eff$ do change.
\par
In appendix B we derive an expression for the energy in general \lag\
theories, which, specialized to the case of MOND, reads
$$H=\lk-2\ao\pd{\lk}{\ao}+\phi+\dot Q,  \eqn{\xxix} $$
where
$$Q=-\sum\_{i=2}\&{N}\sum\_{m=1}\&{i-1}(-1)\&{m-1}(i-m)\rder{i-m}\cdot
\lpd{i}{m-1}.   \eqn{\xxx} $$
This has an immediate corollary: for any solution of the \eom\
 on which
$Q$ does not grow as fast as $t$, at $t\rar\pm\infty$, such as on
periodic, and presumably other bounded
trajectories, we can write
$$H=\langle H\rangle=\langle\lk\rangle
-2\ao\pd{\langle\lk\rangle}{\ao}+\langle\phi\rangle.   \eqn{\ggi}$$
Combined with the virial relation\wxvi this gives
$$\langle H\_{k}\rangle=\oot\c\langle\vr\cdot\grad\phi\rangle,
\eqn{\bnbn} $$
where $\c\equiv(1+2\hat\sk)/(1+\hat\sk/2)$. In the Newtonian limit $\c=1$;
in general $\c<4$; for circular orbits in the MOND limit $\c=2$.
\par
If the theory is Galilei
 invariant and is local or weakly non-local,
$\lk$ may be written (up to a total time
derivative)
 in the form
$$\lk=\oot\a v\&{2}+\tilde\lk(\ao,\rder{2},\rder{3},...),  \eqn{\xxii}$$
where $\a$ is a number, and $\tilde\lk$
 depends only on the acceleration
and higher derivatives of $\vrt$. We show this in appendix A.
Correspondence with Newtonian dynamics requires that $\a=1$, and that
$\tilde\lk$ vanish for $\ao=0$
(or reduce to a time
derivative, which we can eliminate from the outset).
On the other hand, the MOND limit requires $\a=0$.
Thus, {\it  Candidate, Galilei-invariant
theories for MOND,
 that are derivable from an action, must be strongly non-local}.
This is the basis for our statement that eq.\form ,
which is local, and Galilei invariant,
 cannot be derived from an action.

Higher-derivative \lag\ theories that are not strongly non-local
 may be cast in a Hamiltonian form
as described by Ostrogradski\REF{\ostro}{M. Ostrogradski, Mem. Act.
St. Petersbourg {\bf VI4,} 385 (1850).}\refend.
Very succinctly, one defines $N$ coordinates, the first of which
is $\vec q\_{1}\equiv\vr$; the rest are defined recursively by
$\vec q\_{m}=\deriv
{\vec q\_{m-1}}{t},~~m=2,...,N$. There are $N$ momenta, which are
those defined by eq.\xxvii .
The Hamiltonian is $H$, as defined in \xxiv ,
 in which the derivatives of
$\vrt$ have been expressed by the $2N$ coordinates and momenta,
taken to be independent. The Hamilton equations of the usual form
then give the \eom , and, to boot,
 the above relations between the coordinates.
The Hamiltonian formalism is particularly
 useful when one endeavors to quantize
the theory; we shall have little use for it here.
\vskip .15truein
\bk
{\it SOME HEURISTIC MODEL ACTIONS}
\par
Before embarking on general issues concerning properties of the
 solutions, we describe a few examples of theories on which we
shall demonstrate various points.
\par
Consider first a class of kinetic \lag s of the
 form
$$\lk=\oot\vv f(\aos \dleft g\&{-1} \dright)\vv.   \eqn{\xxxi}$$
Here,
 $\dleft$ and $\dright$ are the time derivative operators acting
to their left and right respectively,
and $g$ is a function of the derivatives of $\vrt$ from the second
upward, having the
right dimensions
 ($l\&{2}/t\&{6}$)
 to make the argument of $f$ dimensionless.
 We must have $f(0)=1$ for Newtonian correspondence; furthermore,
as stated in point (iii) in the beginning of this section,
for $\lk$ to approach the Newtonian limit for all trajectories,
 $f(z)$ has to be non-singular at $z=0$ in the complex plane.
If it has a Taylor expansion, with a non-zero radius
of convergence:
 $$f(z)=1+\sum\_{m=1}\&{\infty}b\_{m}z\&{m},  \eqn{\xxxii}$$
then
$$\lk=\oot v\&{2}+\oot b\_{1}\aos g\&{-1}a\&{2}+\oot b\_{2}\ao\&{4}
\left\({d(g\&{-1}\va)\over dt}\right\)^2+O(\ao\&{6}),  \eqn{\xxxiii}$$
which has the form \xxii .
Clearly, $\lk$ may not be represented by this expansion for all
values of $\ao$, lest it not be strongly non-local,
 and not have a MOND
limit.
In the next section we derive an exact expression
for the rotation curve in this class of theories, and show that to obtain
asymptotically flat rotation curves we must have
$f(x)\propto x\&{-1/2}$ for $x\rar\infty$ ($x$ real).
For example
 $f(z)=(1+z)\&{-1/2}$
 satisfies all the above requirements; it has a finite radius
of convergence.
\par
The \lag \xxxi is Galilei invariant: the required Newtonian
limit implies that
$f(\aos \dleft g\&{-1}\dright)$ acting on a constant vector $\vv\_{0}$,
to its left or right,
 gives
$\vv\_{0}$ itself; so, under a transformation
 $\vrt\rar\vrt+\vv\_{0}t+\vr_0$, $\lk$ changes by
 $v\_{0}\&{2}/2+\vv\_{0}\cdot\vv(t)$, which is a total time derivative.
\par
Second, consider \lag s of the form
$$\lk=\oot\vv f\(\aos (\dleft h \dright)\&{-1}\)\vv.
  \eqn{\qqqi}$$
Here, Galilei invariance is achieved because of the required MOND
limit $f(\infty)=0$.
\par
We consider also
actions of the form \x, where $\lk$ depends also on parameters, $g\_{i}$,
that are
 functionals of the whole
trajectory (and not functions of the local values of $\vrt$, and
its derivatives).
 Take, for instance $\lk$ of the form \xxxi \(or \qqqi\)
with $g$ (or $h$) of the form
$$g=\langle G\rangle\equiv\aver{G},    \eqn{\zzi}$$
where $G$ is a function of the derivatives of $\vrt$.
The action is not, strictly speaking, \lag ; we keep that epithet
but call such \lag s compound.
When all the $g\_{i}$s are of the form $g\_{i}=\langle G\_{i}\rangle$,
with local $G\_{i}$s,
the \eom\ has the same form as the Euler-Lagrange equation, with $\lk$
replaced by
$$\mk=\lk+\sum\_{i}G\_{i}\pd{\lk}{g\_{i}}.  \eqn{\pppi} $$
\par
In an even wieldier class of (non-\lag)
 theories we define the  kinetic action $\sk$ as a solution of the
equation
$$\sk=\oot\aver{\vv\cdot
 F\left({\sk \DD\over\aos}\right)\vv}. \eqn{\zzii} $$
The integrand is defined when $\vvt$ can be written as a sum of
 exponentials $exp(\a t)$, with arbitrary complex $\a$:
$e\&{\a t} F\left({\sk \DD\over\aos}\right)e\&{\b t}=
F\left({\sk \a\b\over\aos}\right)e\&{(\a+\b)t}$.
In the Newtonian limit $F(x)\xlimin 1$. To get the correct MOND limit
for circular orbits,
we must have $F(x)=(x/\eta)\&{1/3}$ for $x\ll 1$ (with our normalization
of $\ao$, $\eta=9/8$).
 $F(z)$ is analytic at infinity, according
to our general requirement, but is non-analytic at $z=0$.
This theory is ``almost'' linear, whence come its heuristic value.
\par
Note that only trajectories that can be expanded in real frequencies are
relevant here: For an infinite-time world line
there is no finite-$\sk$ solution
to eq.\zzii, if there are complex-frequency components to $\vrt$.
Such orbits necessarily approach the Newtonian regime, for large
(positive or negative) times,
 but they are
irrelevant there as candidate solutions of the \eom .
\par
The \eom\ is
$$(1+\hat\sk/2) F\left(-{\sk D^2\over\aos}\right)D^2\vrt
=-\grad\phi \eqn{\zziii} $$
with $\hat\sk\equiv-\pd{ln~\sk}{ln~\ao}$.
\par
To obtain eq.\zziii\ we write
$$\d\sk={1\over 2T}\int dt~
\left\(\deriv{\d\vrt}{t}\cdot F\left({\sk \DD\over\aos}\right)\vvt+
\vvt\cdot F\left({\sk \DD\over\aos}\right)\deriv{\d\vrt}{t}\right\)
\hfill$$
$$\hfill +\d\sk{1\over 2T}\int dt~\vvt
 F'\left({\sk \DD\over\aos}\right){\DD\over\aos}\vvt.
 \eqn{\zziv} $$
The two terms in the first integral are equal, as $\DD$
acts symmetrically. The second integral can be written in
terms of $\hat\sk$, leading to
$$\d\sk=(1+\hat\sk/2){1\over T}\int dt~
\deriv{\d\vrt}{t}\cdot
 F\left({\sk \DD\over\aos}\right)\vvt.  \eqn{\zzv}$$
Now, when $\vrt$ and $\d\vrt$ can be expanded in Fourier components
with real frequencies, the last integral equals
$-\int dt~\d\vrt\cdot F\left(-{\sk D^2\over\aos}\right)D^2\vrt$, {\it up
 to end terms} \(when $F(z)$ is analytic at $z=0$ this is seen by
integration by parts; when we go via Fourier space the end terms are
lost\), from which follows eq.\zziii.
\par
To insure
 Galilei invariance, $\lk$ must actually be invariant,
not just incremented by a time derivative; $\sk$ itself
must be invariant--not just changed by end terms--as can be seen
from the \eom . This, in turn, is insured by the required MOND limit
$\(F(0)=0\)$.
(Hence the choice of $\dleft\dright$ in the argument of $F$, in the
definition of $\sk$, is essential. A choice of $-D^2$ instead would
have given an expression that differs by an ends term, and that is
not exactly invariant under Galilei transformations.)
We want $\sk$ to
be determined uniquely by eq.\zzii. If this has many solutions, for a
given trajectory,
we need an added prescription to choose from among them.
The obvious desideratum that $\sk$ go to its Newtonian value, for
 that trajectory, when
$\ao\rar 0$ may do (see next subsection);
 it eliminates $\sk=0$, which is
always a solution.
Note that the theory is time-reversal symmetric (if $\sk$ is a solution
so is its value for the time reversed orbit, and the two have the same
Newtonian limit).
\par
Finally, we bring for comparison a local \lag\ theory with
$$\lk=\oot\vsq f(\aooas);   \eqn{\pppii} $$
obviously it not Galilei invariant.
\vskip .15truein
\bk
{\it THE NATURE OF THE SOLUTIONS OF NONLOCAL THEORIES}
\par

Higher-derivative theories are notoriously problematic. Pais and
Uhlenbeck\REF{\pu}{A. Pais and G.E. Uhlenbeck, Phys. Rev.
{\bf 79}, 145 (1950).}\refend, and more recently,
Ja\'en et al.\REF{\jlm}
{X. Ja\'en, J. Llosa,
 and A. Molina, Phys. Rev. {\bf D 34}, 2302 (1986).}\refend,
 Eliezer and Woodard\REF{\ew}
{D.A. Eliezer and R.P. Woodard, Nucl. Phys.
 {\bf B325}, 389 (1989).}\refend, and Simon
\REF{\simon}{J.Z. Simon, Phys. Rev. {\bf D 41}, 3720 (1990).}\refend~
review these problems in \lag\ theories,
describe possible solutions, and give further
references.
Very succinctly, the problems are: (i) The energy
as defined in eq.\xxiv\ is not bound from below: there are solutions
of the \eom\ with arbitrarily small (large negative) energy.
(ii) There appear runaway solutions of the \eom, on which the velocity,
acceleration, etc. blow up at $t\rar\infty$. In linear theories the
divergence is exponential.
 (iii) When $N$ is the
highest derivative on which the \lag\ depends, the \eom\ is of order
$2N$, and require the specification of that many initial conditions.
While this may not be a matter-of-principle problem, it is disconcerting.
There is a connection between these three problems, and they can usually
be eliminated in one fell swoop, by adding appropriate constraints
to a finite order theory, or
by culling out the majority of
the solutions of the \eom\ that are not analytic in some expansion
parameter\refesh{\jlm,\ew,\simon}.
\par
 We can see the relation
 between the runaway character of a solution and the energy
from eq.\ggi: When the solution is not runaway, the assumption underlying
this relation is valid, and then we see that $H$ is bound from below,
since $\lk$, and$-\pd{\lk}{\ao}$ are positive definite.
\par
When the theory is of infinite order, we may expect exacerbation of
the above problems.
For instance, one may have to dictate an infinite number of initial
conditions, and this may signify a total loss of the
initial-value formulation\refesh{\ew}.
\par
Can MOND-oriented theories avoid these daunting problems?
As has been emphasized by Eliezer and Woodard\refesh{\ew} a strongly
non-local theory (in our terminology) is not a higher-derivative theory,
and need not suffer from any of the problems endemic to such theories.
We saw that Galilei-invariant MOND theories must be strongly non-local
and may well be trouble free.
Theories that emerge as effective approximation to healthy theories,
by an elimination of some degrees of freedom, are obviously as healthy
as their progenitors. As explained in \S\ II
we envisage MOND to come about in just this way.
\par
We now demonstrate that the theory whose action is defined by
eq.\zzii, with a (solvable) harmonic-oscillator potential
 $\phi=\oot\omosq r\&{2}$,
is free of the above problems.
As explained below eq.\zzii, only trajectories that can be expanded
in Fourier components with real frequencies have a chance of solving
the \eom\zziii. From this equation we also see that
 every frequency $\om$ that appears in the expansion must
satisfy
$$(1+\hat\sk/2) F\left({\sk \om\&{2}\over\aos}\right)\om\&{2}
=\omosq. \eqn{\hhiii} $$
We can choose $F$ to give a unique (positive) solution
for $\omega\&{2}$, given $\sk$. The physical trajectories have thus
a single frequency, and can be written as
$$\vrt={1\over\sqrt{2}}(\vr\_{0}e\&{i\omega t}
+\vr\&{*}\_{0}e\&{-i\omega t}). \eqn{\iv} $$
The defining equation of $\sk$, \zzii,
 which supplements the \eom, reads here
$$\sk=\oot\omega\&{2}r\_{0}\&{2}
 F\left({\sk \omega\&{2}\over\aos}\right), \eqn{\hhii} $$
where $r\_{0}\&{2}\equiv\vr\_{0}\cdot\vr\&{*}\_{0}$.
We can also write from eqs.\hhiii\hhii
$$(1+\hat\sk/2) \sk=\oot r\_{0}\&{2}\omosq.  \eqn{\hhv} $$
Take for instance
$$F(x)=(1+\eta x\&{-1})\&{-1/3},  \eqn{\hhvi} $$
which has the desired Newtonian and MOND limits (with $\eta=9/8$).
It is easy to ascertain that $Re\(\vr\_{0}\)$
and $Im\(\vr\_{0}\)$
determine $\omega\&{2}$ (and $\sk$) uniquely.
\par
Thus, there are no runaway solutions, and the physical trajectories are
determined by two initial conditions, just as for a Newtonian
harmonic oscillator. Here too the motion is harmonic, but the frequency
depends on the amplitude. For example, in the deep MOND case--i.e.,
when $\omosq r\_{0}\ll\ao$, we have $\hat\sk\approx 1$,
and $F(x)\approx(x/\eta)\&{1/3}$, and $\omega$ is determined by
$$\om\&{2}\approx\omosq
\left({\om\_{0}\&{4}r\_{0}\&{2}\over\aos}\right)\&{-1/4}\gg\omosq.
 \eqn{\hhvii} $$
A circular orbit is a special case which we shall discuss below.
\par
The case of an anisotropic harmonic potential, which is as easy
to solve, is also instructive. The motion is still harmonic with a
unique frequency in each of the principal directions of the potential,
the equation of motion now being
$$(1+\hat\sk/2) F\left({\sk \om\_{m}\&{2}\over\aos}\right)\om\_{m}\&{2}
=\omosqm,~~~m=1,2,3. \eqn{\hhvii} $$
with
$$(1+\hat\sk/2) \sk=\sum\_{m=1}\&{3}
\vert r\_{0m}\vert\&{2}\omosqm.
  \eqn{\hhviii} $$
Given the three amplitudes and phases encoded in the complex
amplitudes $r\_{0m}$ one determines $\sk$ and the three frequencies
$\om\_{m}$; each depends not only on its own amplitude;
they are coupled through
$\sk$.
\par
In general, exponentially exploding solutions are precluded in MOND,
because they posses derivatives that all blow up at
 infinite time, and this
necessarily takes us to the Newtonian regime where
they are not physical trajectories. In finite-order MOND theories
even trajectories that runaway as a high enough power of $t$ are
precluded.
 This, in itself, does not
solve any of the problems listed above. We saw that the criterion for
a non-runaway solution is that $Q$ of eq.\xxx not grow as fast as $t$
at $\pm\infty$.
 Take, for instance,
the local MOND \lag\ given by eq.\pppii. Clearly, the \eom\ cannot have
solutions for which $a$ increases indefinitely at $t\rar\infty$; still,
the theory has all the usual illnesses: The runaway solutions are
 such that $v\propto t\&{1/3}$ for $t\rar\infty$, so $a\rar 0$,
but the $\dot Q$ term in the expression for $H$ is $\propto v\&{3}$,
enough to qualify as a runaway.

\vskip .3truein
{\bf IV. CIRCULAR ORBITS: DYNAMICS OF AXISYMMETRIC DISK GALAXIES}
\par
No other observation is as central
 to the discussion of the mass discrepancy
in galactic systems, and as clear-cut,
 as the measurement and analysis of rotation curves
of disk galaxies. Fortunately, the relevant trajectories are particularly
amenable to treatment in the framework that we consider here.
These are circular,
constant-speed orbits (the latter epithet will be implicit hereafter) in
the mid-plane of an axisymmetric, and plane symmetric potential field,
 $\phi$.
\vskip .15truein
\bk
{\it THE EQUATION OF MOTION--ROTATION CURVES}
\par
We now show that, under the MOND assumption, and
 for a general action of the type given by eqs.\vi-\iii, the
speed, $v$, on a circular orbit, in the mid-plane of a potential with the
symmetry of a disk,
 depends on the orbital radius, $r$,
through a relation of the form
$$\m(a/\ao)a=\pd{\phi}{r},   \eqn{\rc}$$
where $\m(x)$ depends on the specific kinetic action,
 and $a=v\&{2}/r$.
Thus, eq.\form gives the exact expression for the rotation curves,
and existing rotation-curve analyses, which
have all employed this
 expression\refesh{\mib,\kent,\mirc,\bbs},
 are valid in the present framework.
\par
For consistency we first ascertain that, for a circular orbit, the
functional derivative, $\A$, in the \eom , is a radial vector
of a constant magnitude. This is a property of the kinetic action
alone (irrespective of the potential) and is necessary if circular
orbits  are to be solutions of the \eom\
in the mid-plane.
This property follows from the assumed
 translation and rotation invariances of the
kinetic action.
Under a infinitesimal translation $\vec r\rar\vec r+\vec \epsilon$,
$\sk$ changes by $\vec\epsilon\cdot\int\A~dt$,
 according to eq.\vi\ for the variation of $\sk$
\(the ends term vanishes as $\vrt$ and $\d\vrt$ are periodic\).
This must vanish for
arbitrary $\vec\epsilon$, so $\int\A~dt=0$. From the symmetry of the
orbit, the component of $\A$ perpendicular to the orbit must be constant
and hence must vanish.
Similarly, the invariance of $\sk$ under
 a rotation about an arbitrary
 axis $\vec n$ ($\d\vr\propto\vec n\times\vr$)
 implies that
$\int\A\times\vrt~dt=0.$
 Thus, $\A$ cannot have a tangential component either, and is a
radial vector of constant magnitude, as required.
The action underpinnings is essential;
 it is easy to write an \eom\vii
in which $\A$ has the above
symmetries but is not radial for circular orbits.
(But TR is then not a symmetry.)
\par
The magnitude of $\A$ depends only on the
parameters of the orbit, to wit $v$, and $r$. Because $A/a$ is
dimensionless, it must, by the MOND assumption, be of the form
$\m(a/\ao)$, leading to
$$\A=\m(a/\ao)\va,  \eqn{\ama}$$
and
 to eq.\rc , in turn. For there to be a circular-orbit
 solution
for every positive value of $\pd{\phi}{r}$, $x\m(x)$
must take all positive values, and $\A$ must be directed inward.
Uniqueness requires that $x\m(x)$ is monotonic.
As a corollary we get that any kinetic action that is
rotation and translation invariant necessarily
 gives the Newtonian relation
for circular orbits, if it does not contain any dimensional constant.
\par
There exists a simple and useful expression for $\m(x)$ in terms
of the values of the action for circular orbits:
$$\m(a/\ao)=2v^{-2}\skc(1+\oot\hat\skc),  \eqn{\formu}$$
where $\skc$ is the action calculated for a circular orbit,
 and $\hat\skc=-\pd{ln~\skc}{ln~\ao}$.
This relation enables us to circumvent
the calculation of functional derivatives.
\par
Equation\formu follows from the general expression \wxv for
$\langle\A\cdot\vr\rangle$--which holds for any periodic orbit--with
 eq.\ama , and the equality $\va\cdot\vr=-v\&{2}$.
\par
With our normalization, $\sk$ has the dimensions of velocity square;
so, $\skc$ must be of the form
$$\skc=\oot v\&{2}\l(a/\ao).  \eqn{\icirc}  $$
(Any functional of the orbit reduces to a function of
$v$, $r$, and $\ao$, for a circular orbit.
 Thus, $\skc/v\&{2}$ is a dimensionless function of
these, and must be of the form $\oot \l(v\&{2}/r\ao).$)
Hence,
$$\m(x)=\l(x)+\oot x\deriv{\l}{x}=\l(x)
\(1+\hat\l(x)/2\),
\eqn{\mula}  $$
where $\hat\l(x)\equiv dln~\l/dln~x$.
\par
Only the time-reversal-symmetric part of $\sk$ contributes to $\skc$,
and hence to $\m$,
because, for a circular orbit, time reversal is equivalent to some
$\pi$-rotation about an axis in the orbital plane, under which $\sk$ is
invariant.
\par
When the action is derivable from a \lag\ ,
 $\lk$--simple or compound, local or not--that does not depend
on the time explicitly,
 one has
$\skc=\lkc $, where $\lkc$ is the constant value of the
 \lag\ on the circular orbit.
Then
$$\m(a/\ao)=2v^{-2}\lkc(1+\oot\hat\lkc),  \eqn{\xvi}$$
with $\hat\lkc=-\pd{ln~\lkc}{ln~\ao}$.
\par
Consider, for instance, \lag\ actions with $\lk$ of the form \xxxi .
The function $g$ is constant on a circular orbit, and, in fact,
must be proportional to $r\&{-4}v\&{6}$, on dimensional grounds.
 The  proportionality constant may be absorbed into $\ao$.
The operator $\dleft\dright$, when sandwiched between any two derivatives
of $\vrc(t)$ (or
 $-D^2$, when acting on any of these),
is just the multiplication by $\omega^{2}=v\&{2}/r^2$. We then find that
$$\lkc=\oot v\&{2}f(\aooas),   \eqn{\xviii} $$
from which results the expression
$$\m(a/\ao)=
f(x)\(1+\hat f(x)\)\vert\_{x=\aooas},  \eqn{\xix}$$
where
$\hat f(x)\equiv -dln~f/ dln~x$.
\par
Since $\m$ does not depend on the choice of $g$, many theories
give the same rotation curve for a given mass distribution. For similar
reasons, \lag s of the form\qqqi\ give the same rotation curve (with
the same choice of $f$), although they are very different
theories. The same is true of the compound \lag s discussed below
eq.\qqqi, and of the local \lag\ given by eq.\pppii.
\par
For actions of the form \zzii,
we find from the general expression for $\m(x)$ eq.\formu
\(or directly from the \eom \zziii\)
 that $\m(x)=\(1+\oot\hat\l(x)\)\l(x)$, where $\l(x)$ is the solution of
 $\l(x)=F\(x^2\l(x)/2\)$ with the correct Newtonian limit; i.e.,
the solution, for $\a=0$, of
 $\l\_{\a}(x)=F\(x^2\a\l\_{\a}(x)/2\)$ that satisfies
 $\l\_{\a}(x)\rar 1$ for $\a\rar\infty$.
\par
Clearly, we can add to the \lag\ a time-reversal-antisymmetric part
without affecting the rotation curve. The same is true of a more general
type of terms that vanish on circular orbits
 \(e.g. $\propto(\vv\cdot\va)^2$\).
\par
The fact that the rotation curve, for a given mass distribution
depends only on the value of the action for circular orbits
is gratifying in that it allows us to test this
general class of theories by rotation-curve analysis, without having
to know the exact theory \(by adopting a best value for $\ao$, and
a reasonable form for $\m(x)$\). This, however, leaves us bereft
of an important observational guide in discriminating between
 theories within the class.
\vskip .15truein
\bk
{\it THE ENERGY AND ANGULAR MOMENTUM }
\par
When the action is derivable from a \lag , and energy
and angular momentum are defined--in the sense that the formal sums
that express them converge, or may be understood as analytic
 continuations--we find simple expressions for these
quantities on circular orbits.
\par
Because scalar functions are constant on circular
 orbits, $\dot Q$ in expression\xxix for the energy
vanishes and we have
$$H_c=\lkc-2\ao\pd{\lkc}{\ao}+\phi.  \eqn{\xl} $$
\par
In terms of $\l(x)$ of eq.\icirc, which now describes
the dependence of $\lkc$ on $a/\ao$,
we can write
$$H_c=\oot v\&{2}\n(a/\ao)+\phi,  \eqn{\xli} $$
where
$$\n(x)=\l(x)\(1+2\hat\l(x)\)=\m(x)\kappa(x),~~~
\kappa(x)={1+2\hat\l(x)\over
1+\hat\l(x)/2}. \eqn{\xlii}  $$
In the deep MOND limit, where $\m(x)\approx x$, and $\hat\l(x)\approx 1$,
$\n(x)\approx 2x$.
\par
Putting together eqs.\xli ,\xlii , and eq.\rc\ $\(\m(a/\ao)a=\phi '\)$
 which governs circular
orbits in a disk potential $\phi$ (taken to vanish at infinity),
 we can write for such a motion
$$H_c=\oot\m(a/\ao)v\&{2}(\kappa+2/\hat\phi),  \eqn{\xlviii}$$
where $\hat\phi\equiv dln~\phi/dln~r$.
In Newtonian dynamics ($\m=1,~\kappa=1$)
 there are no
 negative-energy circular orbits for potentials with $\hat\phi\le -2$.
It is interesting to note that in the strict MOND
limit we have $\kappa\rar 2$,
and the limiting potential
approaches the Newtonian or Coulomb potential, with $\hat\phi\rar -1$.
\par
Coming now to the angular momentum, we
note that, for a circular orbit, vectors in the orbital plane
 \($\vrt$ and its
derivatives, derivative of $\lk$ with respect to those, etc.\) are
radial or tangential, according to whether
they are even or odd under time reversal, respectively-- because, for
such orbits,
 time reversal is equivalent to some $\pi$ rotation about a diameter
of the orbit. We see then
from expression \xxva\ for $\vec J$ that only the time-reversal-symmetric
part of $\lk$ contributes. Furthermore, the summand in eq.\xxva can now
be written (after successively shifting time derivatives from the
 second factor to the first) as $\rder{m-1}\times\pd{\lk}{\rder{m}}$.
Hence, we can write for a circular orbit
$$\vec J_c=\sum\_{m=1}\&{N}m\rder{m-1}\times\lp{m}. \eqn{\xliv}$$
Since $\rder{m-1}=-\omega^{-2}\vec\omega\times\rder{m}$,
with $\vec\omega=r^{-2}\vr\times\vv$, we have
$$\vec J_c=\omega^{-2}\vec\omega
\sum\_{m=1}\&{N}m\rder{m}\cdot\lp{m}. \eqn{\xlv}$$
Finally, from expression (B-4) in appendix B, for this sum, we get
$$\vec J_c={2\over \omega^2}\vec\omega\left(\lkc-\ao\pd{\lkc}
{\ao}\right). \eqn{\xlvi} $$
In terms of $\l(x)$ of eq.\icirc
we have
$$\vec J_c=\eta(a/\ao)\vr\times\vv,~~~~~
\eta(x)=\l(x)\(1+\hat\l(x)\)=\m(x){1+\hat\l(x)\over 1+\hat\l(x)/2}.
 \eqn{\xlvii} $$
So, in the very MOND limit $J_c\rar 4v\&{3}/3\ao$, and, interestingly,
depends only on $v$.
\par
In modifications of gravity, the expression for the angular momentum
is the same as in conventional dynamics; so, the adiabatic invariance of
$\vec J$ has rather different consequences in the two classes of theories
based, respectively, on modifying gravity, and modifying the law of
motion.

\vskip .3truein
{\bf V. MODIFIED DYNAMICS AT LOW FREQUENCIES}
\par
Notwithstanding the phenomenological success of the MOND scheme, it is
of interest to consider modifications that hinge on other system
 properties.
 After all, galactic systems differ from laboratory
and planetary systems in attributes other than the typical accelerations
in them.
Here we examine modifications
that hinge on the time scale involved. More accurately, we
have in mind theories that involve a constant, $\Om$,
 with the dimensions of frequency. In analogy with the above discussion
 of MOND, the \eom\ is
of the form
$$\A\(\vrt,\Om\)=-\grad\phi.  \eqn{\tto} $$
  From dimensional considerations we find that
$$\A\(\vrt,\Om\)=\Om^2\A\(\vr(\Om t),1\),  \eqn{\ttii}$$
and that
$$\A\(\l\vrt,\Om\)=\l\A\(\vrt,\Om\),  \eqn{\ttiii}$$
Thus, $\A$ is necessarily homogeneous in $\vrt$, and there is nothing
to prevent us from having $\A$ that is linear in $\vrt$.
Theories with a linear kinetic term
 are rather more amenable to analysis; in particular their \eom\ is
easy to derive from the action, when they are thus derivable;
 hence their
heuristic value. Henceforth we shall limit ourselves to such
\eoms ,
whose general form is
$$f(D/\Om)D^2\vrt=-\grad\phi.   \eqn{\ttiv}$$
\par
In the present context the appropriate Newtonian limit is $\Om\rar 0$, if
we want approximate Newtonian behavior for laboratory phenomena,
where the typical frequencies would be much higher than $\Om$.
Thus,
$$f(x)\xlimin 1.  \eqn{\ttv}$$
Galilei invariance of $\A$
 is assured if $f(x)$ is not
singular at $x=0$.
\par
If we wish to follow this course beyond heuristics, and
propound such theories as potential alternatives to dark matter,
we have to note the following:
For circular orbits in the mid-plane of a disk potential
to be solutions of the \eom\ $f(x)$ must be an even function of $x$:
$f(x)=h(-x\&{2})$.
Then, eq.\ttiv gives for such orbits
$$h\((v/r\Om)\&{2}\)v\&{2}/r=\deriv{\phi}{r}. \eqn{\ttvii} $$
Flat rotation curves for disk galaxies result if
$$h(y)\approx y\&{1/2},~~~~for~~y\ll 1. \eqn{\ttvi}$$
The asymptotic
rotational speed, $\vin$, is then related to the total galactic mass by
$\vin\&{3}=MG\Om$. This is at some odds with the observed Infra-red
Tully-Fisher relation ($\vin\&{4}\propto L\_{IR}$),
 if the IR luminosity of a galaxy
 is proportional to its mass (see point 2 in
\S\ I).
The value of $\Om$ would have to be about $10\&{-3}\h0$
to account for the mass discrepancy in galaxies (this value
is approximately
the orbital frequency of the sun in the Galaxy). Typical dynamical
time scales in clusters of galaxies are much larger than in galaxies,
so we would predict rather higher discrepancies in clusters.
Not much of an Oort discrepancy is expected in the solar neighborhood,
as the frequencies of the perpendicular motion are at least a few times
higher than $\Om$, and due to the linearity of the kinetic term
 there is no direct effect of the low-frequency
orbital motion on the perpendicular motion (there is only a shift
in the equilibrium state for the oscillations).
This linearity also assures the desired non-Newtonian
 center-of-mass motion of
composite systems (stars, clouds, or atoms in galaxies, etc.)
even though their components
rattle inside them at high frequencies.
\par
Phenomena in the laboratory or solar system are typified by frequencies,
$\omega$, much higher than $\Om$, and if the correction to Newtonian
 laws, in the limit $\omega\gg\Om$, is of order $(\Om/\omega)^2$--as
expected if $f(x)$ is even in its argument--it would not be
 detectable with present experimental capacity.
\par
The evenness of $f$ in its argument is tantamount to time-reversal (TR)
symmetry of the \eom , and also to its derivability
from a \lag\ action with
$$\lk=\oot\vv(t)h\left({\dleft\dright\over\Om^2}\right)\vv(t).
 \eqn{\ttvix} $$
(Only the time-symmetric part of the \lag\ affects the \eom, which is
thus necessarily TR symmetric.)
\par
Pais and Uhlenbeck\refesh{\pu}
have discussed in detail the properties of
a class of theories with \lag s equivalent to that of eq.\ttvix,
with $h(z)$ an entire (integral) function.
They show that such theories suffer from the problems discussed in
\S\ III.
Our insistence on the particular Newtonian
limit \ttv\ (which they are not concerned with)
puts us outside the province of their analysis:
An entire function has an
essential singularity at infinity (barring the conventional case of
a constant $f$),
 and thus cannot have the Newtonian limit
as required in eq.\ttv .
(For a valid Newtonian limit, $f(z)$ must go to
1, as $z\rar\infty$, in all directions in the complex plane, if the
\lag\ is to approach the Newtonian \lag\ for all trajectories. For
 example, harmonic trajectories correspond to an imaginary argument
of $f$; time-exponential trajectories correspond to real ones, etc..)
Our theory must then be non-local, indeed strongly
non-local. In the present context this follows only from the required
Newtonian limit; in the case of MOND, Galilei invariance, and the MOND
limit were also brought to bear on the analogous conclusion.
As we explain in \S\ III, the fact that our theory must not be entire in
the time-derivative operator is a blessing: only such theories
have a chance of being free of the difficulties that characterize
 higher-derivative theories.
\vskip .3truein
{\bf VI. DISCUSSION}
\par
Examining modified laws of motion
in the framework of the MOND scheme, we find that
  if the underlying kinetic
action has the correct Newtonian limit, has the phenomenologically
inspired MOND limit, and is Galilei invariant, the theory must be
strongly non-local. It is this class of theories that we have
concentrated on. They are not infinite-order
 limits of higher-derivative theories, which are known to have severe
problems. We have
 demonstrated that MOND non-local theories may be formulated
that are free of these problems. An attractive possibility is that
MOND results as a nonrelativistic, small-scale
expression of a
fundamental theory by which inertia is a vestige of the interaction
of a body with ``the rest of the Universe'', in the spirit of Mach's
principle, or with an agent such as vacuum fields.
 The approximate equality of the acceleration ``constant''
of MOND, $\ao$, and $c\h0$, may be a witness. Such effective theories
are usually strongly non-local.
\par
We are not able yet to pinpoint a specific effective theory.
 The study of rotation curves--which provides
the most detailed and clear-cut evidence for a mass discrepancy--is
a strong test that is passed by all theories of the genre we
consider here; but, it can only constrain the values of the kinetic
 action on circular orbits. Further constraints could
come from the detection
of mass discrepancies in light bending.
Beyond these,
existing observations of the mass discrepancy in galactic systems
are hardly helpful in discerning
a superior candidate theory.
Additional discriminating power might come from the study of
systems with non-circular motions, such as elliptical galaxies and
clusters of galaxies, because theories that predict the same circular
motion usually disagree on non-circular trajectories.
\par
 Also potentially helpful is the study of
 a possible mass discrepancy manifested by the motion
perpendicular to the plane in disk galaxies, which is a small
 perturbation on the circular motion. There may be such a mass
 discrepancy observed in the solar neighborhood, in the Milky Way
(the Oort discrepancy), but the issue is still moot. Different theories
of the present class predict different degrees of such a discrepancy.
For example, it was pointed out\refesh{\mib} that eq.\form\ predicts
a discrepancy of magnitude $1/\m(a\_{c}/\ao)$ where $a\_{c}$ is
the acceleration in the circular orbit (as long as the acceleration
perpendicular to the disk is small compared with $a\_{c}$).
Analysis of small-perturbation, perpendicular
motion in the theory defined by the action \zzii, gives a different
value for the discrepancy: $1/\((1+\hat\skc/2)F(\skc\om\&{2}/\aos)\)$,
where $\om$ is the frequency of the perpendicular motion. Had
$\om$ been the
frequency of the circular motion (actually $\ll\om$), this would
have given $1/\m(a\_{c}/\ao)$.
\par
The examples we discuss are still formulated
as test-particle dynamics.
 We cannot yet offer an
example of a theory that fully accounts for the correct center-of-mass
motion of composite objects: A composite body (e.g. a star),
 with constituents
whose internal motion is Newtonian should
still undergo a center-of-mass motion similar to that of
 a test particle, which may be well in the MOND regime.
This we deem only a technical obstacle; a feeling that is supported by
the following observation.
We may formulate a MOND action as an integral over the
Fourier transform, $r\_{\om}$,
 of the trajectory, such that, in an extreme
case, different frequencies are decoupled from each other,
and for each frequency the motion depends on the value of
$\om\&{2}r\_{\om}/\ao$. Thus, the internal motions for which this
value is large, obey approximate Newtonian dynamics, while the
center-of-mass
motion, characterized by a small value, is test-particle like.
The theories discussed in \S\ V have this desired property.
 For various reasons we do not endorse
 such total spectral decoupling of the motion in the kinetic action
(as occurs in the Newtonian action).
\par
 The action in eq.\zzii
does couple different frequencies, and also give a partial solution
to the problem: Consider a trajectory
of a constituent of a ``star'' that performs
 ``galactic'' motion with (say one) frequency $\omg$
and frequencies $\omi$ that describe the internal motion.
We see from eq.\zzii that, if the internal velocities are small enough
compared with the galactic one, the value of $\sk$
is determined by the ``galactic'' component, even if
$\omi\gg\omg$ because $F$ saturates at a value of 1 for
large arguments. The condition for that is
$$(v\_{i}/v\_{g})\&{2}\ll\sk\omg\&{2}/\aos. \eqn{\bbbbii} $$
Then, the ``galactic'' motion will be test-particle like, and the
internal motions will depend on the value of
$(\sk/ v\_{i}\&{2})(a\_{i}\&{2}\aos)$, where
$a\_{i}\equiv v\_{i}\omi$.
This theory does not provide a
fully satisfactory solution  because it does not cover the case
with internal velocities that are not smaller than the external one
(motions within the atoms or nuclei are also internal).
It does indicate, however that the solution is within reach.
\par
Another desideratum we have yet to implement is an extension of the
theory to the relativistic regime: a matter-of-principle must,
that is also needed to treat light bending.
We feel that understanding the underpinning of MOND is a prior goal, and
it achievement will surely bring about the understanding of
relativistic MOND.
We note in this connection that for a phenomenon to be both
 relativistic,
and to involve accelerations of order $\ao$ or below, it must also
 involve a length scale of cosmological magnitude.
For instance, for the
 acceleration near the Schwarzschild radius of a mass $M$ to be
at $\ao$ or below we must have this radius no smaller than the Hubble
radius. Similarly, light trajectories in regions where the Newtonian
acceleration is of order $\ao$, has a radius of curvature
of the order of the Hubble radius.
Thus, the relativistic-MOND regime falls in the large-scales regime.
 In light of the surmised cosmological origin of MOND this
means that a relativistic extension of MOND might not simply hinge on
the single acceleration constant $\ao$, which may, in fact, loose its
role altogether.
\par
Like all effective theories, ours must have validity bounds
 that stem from
the approximations underlying the theory. These delimiters are usually
not recognized from within the theory itself; but, sometimes,
inconsistencies or other problems are vestiges of these approximations.
Consider, for instance trajectories with constant velocity;
in all the model actions we have studied the correspond to the
extreme MOND limit in which the action vanishes. Clearly they do not
have a Newtonian limit, which is one of the
axioms of our treatment; we see a hint of a delimiter here.
We should also not be dismayed if our effective (relativistic) theory
fails to account for cosmology; as explained in \S\ I
 we certainly expect a delimiter there.
\par
Finally, we list some differences between
modifying gravity, and modifying inertia,
beyond the fact that the latter pertains to
arbitrary combinations of forces. We saw that
conserved quantities and adiabatic invariants are different in the
two approaches. In gravitational MOND theories,
 any isolated mass produces
a confining gravitational potential: To obtain asymptotically flat
rotation curves we must have an asymptotically logarithmic potential.
It is not clear to us whether there a general answer to the analogue
question in our class of theories (i.e., are there unbound
trajectories about an isolated mass?). The approximation
underlying our theory may, in fact, break down for such unbound orbit,
as indicated above.

\vskip .3truein
\bk
{\bf APPENDIX A}
\par
We derive the general  form of a Galilei-invariant,
local, kinetic \lag ,
 $\lk$ \(which
 can be written as a
 function of $\vrt$ and a finite number of its derivatives\).
We make use of the commutation rules
$$\pd{}{\vr}\ddt{}Q=\ddt{}\pd{}{\vr}Q, \eqno(A-1) $$
and
$$\pd{}{\rder{i}}\ddt{}Q=\ddt{}\pd{}{\rder{i}}Q+\pd{Q}{\rder{i-1}},~~i>0.
 \eqno(A-2) $$
\(These stem from
 $\dot Q=(\vv\cdot\pd{}{\vr}+\va\cdot\pd{}{\vv}+...)Q$.\)
For the \eom\ to be invariant under translations it is necessary
(and sufficient) that $\grad_{r}\lk$ is some time derivative
 $\deriv{\vec Q}{t}$.
 Taking the $r$-curl of this equality we get, with the
aid of (A-1), that
 $\ddt{}\(\vec\nabla_{r}\times\vec Q\)=0$. Now, when a quantity
$F$ depends on a finite number of derivatives of $\vrt$,
and $\ddt{}F=0$ (i.e. $F$ is
constant along any trajectory), then
$F$ must be an absolute constant, independent of any of the variables:
If $F$ depends on $\rder{0},\rder{1},...,\rder{k}$ only, and
$\ddt{}F=\rder{1}\cdot\pd{F}{\rder{0}}+...+
\rder{k+1}\cdot\pd{F}{\rder{k}}=0$, for any choice of $\rder{0},...,
\rder{k+1}$, then, clearly, all the partial derivatives of F must vanish.
(Seen differently by noting that any two {\it finite}
 sets of derivatives are connected by some trajectory, so
$F$ must take the same value on any two sets.)
Thus, $\vec\nabla_{r}\times\vec Q$
is a constant of the theory, and hence must actually vanish,
because of the isotropy of the theory, which forbids the
 construction of a non-vanishing, constant vector.
 Thus, $\vec Q=\vec\nabla_{r}q$ for some $q$, and
hence $\grad_{r}(\lk-\dot q)=0$, and $\lk$ is independent of $\vr$
up to the time derivative $\dot q$, which we can discard.
\par
Take then $\lk$ that depends only on the derivatives of $r(t)$.
For the \eom\ to be Galilei invariant we must have
$\vec\nabla_{v}\lk=\deriv{\vec Q}{t}$ for some $\vec Q$.
The left-hand side does not depend on the variable $\vr$, so
take the derivative with respect to $\vr$ to obtain
$\ddt{}\(\vec\nabla_{r}\otimes\vec Q\)=0$.
As above,
the tensor in parentheses must be a constant, and is dimensionless,
so it must be $\a\d\_{ij}$, with $\a$ a dimensionless constant.
Thus, $\vec Q=\a\vr+\vec U$, where
$\vec U$ is independent of $\vr$. We then have
$\grad_{v}\lk=\a\vv+\deriv{\vec U}{t}$. Taking the $v$-curl (which
now commutes with the $t$-derivative on $\vec U$) we find that
$\vec \nabla_{v}\times\vec U$
is a constant, and hence must vanish, and hence $\vec U$ is a
 $v$-divergence of some $u$ that does not depend on $\vr$.
  From all this we get
$$\grad_{v}(\lk-\oot\a v\&{2}-\dot u)=0.  \eqno(A-3)$$
The expression in parentheses is, thus,
some $\tilde\lk$ that does not depend
on $\vr$ or $\vv$ as stated in eq.\xxii .
\par
If $\lk$ is not of finite order, but is a limit, for $N\rar\infty$,
 of a sequence of \lag s, $\lk\&{N}$, of finite order $N$, then,
 each of the
 $\lk\&{N}~'s$ is of the form \xxii, and, hence, so is $\lk$
(if the $\lk\&{N}~'s$ are not themselves
 Galilei invariant, we can take their
Galilei-symmetrized parts,
 and they too go to $\lk$ for $N\rar\infty$).

\vskip .3truein
\bk
{\bf APPENDIX B}
\par
Here we derive expression\xxix for the energy function, which applies
whenever this function can
be defined.
Starting with the general expression for the energy, eq.\xxiv,
 and shifting time derivatives from the second factor
to the first,
we can write (for $m\ge 2$)
$$\rder{i-m+1}\cdot\lspd{i}{m-1}=\dot q\_{i}+
(-1)\&{m-1}\rder{i}\cdot\lsp{i},
\eqno(B-1)$$
with
$$q\_{i}=-\sum\_{l=1}\&{m-1}(-1)\&{l}\rder{i-m+l}\cdot\lspd{i}{m-l-1}.
\eqno(B-2) $$
Upon taking the double sum in eq\xxiv,
the first term in eq.(B-1) gives $\dot Q$
 with $Q$ given by eq.\xxx\
(after some rearrangement and summation).
 The second term in
eq.(B-1) gives $\sum\_i i\rder{i}\cdot\lsp{i}$, which we now
evaluate.
Imagine that $L$ depends on a number of dimensional constants
$C\_{\ell},~~\ell=1,...,I$,
with dimensions $(length)\&{\a\_{\ell}}
\times (time)\&{\b\_{\ell}}$, respectively.
 The \lag\ itself has dimensions
$(length)\&{2}\times (time)\&{-2}$.
\par
Considering the redefinition of the units of time we have
$$L\(\l\&{\b\_{1}}C\_{1},...,\l\&{\b\_{I}}C\_{I},
\l\&{-1}\rder{1},...,\l\&{-i}\rder{i},...\)=\l^{-2}
L\(C\_{1},...,C\_{I},
\rder{1},...,\rder{i},...\).  \eqno(B-3)$$
Taking the derivative with respect to $\l$ at $\l=1$, we find, as in the
Euler theorem for homogeneous functions,
$$\sum\_{\ell=1}\&{I}\b\_{\ell}C\_{\ell}\pd{L}{C\_{\ell}}
-\sum\_{i=1}\&{N}i\rder{i}\cdot\lsp{i}=-2L.  \eqno(B-4) $$
Putting eq.(B-4) together with the above, we get
$$H=L+\sum\_{\ell=1}\&{I}\b\_{\ell}C\_{\ell}\pd{L}{C\_{\ell}}
+\dot Q.  \eqno(B-5) $$
We separate the \lag \ into kinetic and
 potential parts $L=\lk-\phi$; $\phi$, which is taken to depend only on
$\vr$, must be of the form $Cf(\vr/r\_{0})$, where $C$ has dimensions
of velocity squared, and $r\_{0}$ those of length. Thus the contribution
of $\phi$ to the sum in (B-5) is $2\phi$, and none to $Q$.
We can then also write
$$H=\lk
+\sum\_{\ell=1}\&{I}\b\_{\ell}C\_{\ell}\pd{\lk}{C\_{\ell}}
+\phi+\dot Q,  \eqno(B-6) $$
and specializing to the case of a single constant $\ao$ ($\b=-2$),
 we get eq.\xxix.
\par
Another useful relation is obtained by
considering, instead, redefinition of the units of length, which gives
$$\lk\(\l\&{\a\_{1}}C\_{1},...,\l\&{\a\_{I}}C\_{I},
\l\rder{1},...,\l\rder{i},...\)=\l^2
\lk\(C\_{1},...,C\_{I},
\rder{1},...,\rder{i},...\),  \eqno(B-7)$$
from which follows, as above,
$$\sum\_{\ell=1}\&{I}\a\_{\ell}C\_{\ell}\pd{\lk}{C\_{\ell}}
+\sum\_{i=1}\&{N}\rder{i}\cdot\lp{i}=2\lk.  \eqno(B-8) $$

\endpage
\write2{\bigskip\noindent
  {\tofcfont References\leadtofc\number\count0}\par\smallskip}
\chskipt
\ifodd0{
\rjustline{{\chapfont References}}}
\else{
\ljustline{{\chapfont References}}}\fi
\chskipl
\immediate\closeout 4
\input refc
\endpage
\bye